\documentclass[twocolumn]{aastex62}

\usepackage{amsmath}
\usepackage{amssymb}
\usepackage{empheq}
\usepackage{mathrsfs}
				
\usepackage{amssymb}

\newcommand{\Ham}{\mathcal{H}}
\newcommand{\Kam}{\mathcal{K}}
\newcommand{\G}{\mathcal{G}}
\newcommand{\M}{M_{\star}}

\newcommand{\rr}{\mathbf{r}}
\newcommand{\Poincare}{{Poincar$\acute{\rm{e}}$}}

\newcommand{\be}{\begin{equation}}
\newcommand{\ee}{\end{equation}}

\def\lta{\,\raise 0.3 ex\hbox{$ < $}\kern -0.75 em
 \lower 0.7 ex\hbox{$\sim$}\,}
\def\gta{\,\raise 0.3 ex\hbox{$ > $}\kern -0.75 em
 \lower 0.7 ex\hbox{$\sim$}\,}

\newcommand{\amp}{{\cal S}} 
\newcommand{\diff}{{\cal D}} 

\begin{document}


\title{Dynamics of Planetary Systems Within Star Clusters: Aspects of the Solar System's Early Evolution}

\author{Konstantin Batygin}
\affiliation{Division of Geological and Planetary Sciences California Institute of Technology, Pasadena, CA 91125, USA}

\author{Fred C. Adams}
\affiliation{Physics Department, University of Michigan, Ann Arbor, MI 48109, USA}
\affiliation{Astronomy Department, University of Michigan, Ann Arbor, MI 48109, USA}

\author{Yuri K. Batygin}
\affiliation{Los Alamos National Laboratory, Los Alamos, NM 87545, USA}

\author{Erik A. Petigura}
\affiliation{Department of Physics and Astronomy, University of California, Los Angeles, CA 90095, USA}

\begin{abstract}

Most planetary systems -- including our own -- are born within stellar clusters, where interactions with neighboring stars can help shape the system architecture. This paper develops an orbit-averaged formalism to characterize the cluster's mean-field effects as well as the physics of long-period stellar encounters. Our secular approach allows for an analytic description of the dynamical consequences of the cluster environment on its constituent planetary systems. We analyze special cases of the resulting Hamiltonian, corresponding to eccentricity evolution driven by planar encounters, as well as hyperbolic perturbations upon dissipative disks. We subsequently apply our results to the early evolution of our solar system, where the cluster's collective potential perturbs the solar system's plane, and stellar encounters act to increase the velocity dispersion of the Kuiper belt. Our results are two-fold: first, we find that cluster effects can alter the mean plane of the solar system by $\lesssim1\deg$, and are thus insufficient to explain the $\psi\approx6\deg$ obliquity of the sun. Second, we delineate the extent to which stellar flybys excite the orbital dispersion of the cold classical Kuiper belt, and show that while stellar flybys may grow the cold belt's inclination by the observed amount, the resulting distribution is incompatible with the data. Correspondingly, our calculations place an upper limit on the product of the stellar number density and residence time of the sun in its birth cluster, $\eta\,\tau\lesssim2\times10^4\,$Myr/pc$^3$.

\end{abstract}


\section{Introduction}




Most stars --- and the planetary systems they host --- form within young stellar associations \citep{ladalada,porras}.  An important and ongoing line of inquiry is to understand the manner in which these cluster environments shape the properties of their constituent planetary systems, and thereby further diversify the orbital characteristics of the galactic planetary census. Even the solar system itself exhibits an elaborate and intricate dynamical structure in its distant regions, which is routinely attributed to cluster-induced evolution \citep{MorbidelliLevison2004,Brasser2006}. Although a full explanation for this complexity remains unresolved, the notion that the solar system's birth environment played an important role in sculpting its long-period architecture is rarely contested \citep{adams2010}. The goal of this paper is to explore one aspect of this problem -- the consequences of long-range interactions between planetary systems and individual passing stars, as well as the cumulative gravitational potential of the birth cluster. An understanding of these effects, in turn, provides an important step toward unraveling the age-old question of how planetary systems form and evolve.

Broadly speaking, the theory of planet formation can be divided into two separate themes: the conglomeration of proto-planetary material, and the subsequent dynamical evolution of the planetary system. Although these physical processes are not strictly separable, they nevertheless operate on distinct temporal scales. In particular, assembly of planets is expected to unfold within a geometrically thin disk of gas and dust that dissipates over the course of the first $1-10\,$Myr of the host star's lifetime \citep{armitage}. In contrast, the subsequent dynamical evolution can transpire over much longer timescales, spanning hundreds of Myr \citep{Tsiganis2005,nesvorny}, or even several Gyr \citep{davies,laskar,Batygin2015}.  Moreover, while the process of planet assembly is primarily controlled by {\it local} physics taking place within protoplanetary disks \citep{lambrechts}, dynamical evolution that ensues after a newborn planetary system emerges from its natal nebula can be strongly influenced by its external environment (see \citealt{Hernandez2007,malmberg2007} and references therein).


Various lines of evidence -- including meteoritic enrichment in short-lived radiogenic isotopes, as well as the orbital architecture of the solar system'€s trans-Neptunian region,€" suggest that the Sun itself was born in a cluster of $N\sim10^3-10^4$ stars, where the cluster likely persisted for $\tau \sim10-100$ Myr \citep{adams2010,zwart,brasser2012,pfalzner}. An important consequence of this picture is that planetary systems born within stellar clusters will necessarily experience gravitational perturbations from passing stars. Over the past two decades, extensive numerical investigations of this process have been carried out (see e.g., \citealt{al2001,zwart,malmberg2007,malmberg2011, pfalzner,pfalzner15,liadams15,liadams16}, and references therein). This body of work cumulatively demonstrates how perturbations from stellar encounters and the collective cluster potential can contribute to shaping the orbital architectures of the constituent planetary systems. Nevertheless, a full assessment of these processes is complicated by the diverse nature of stellar birth clusters, which have a wide range of cluster membership size $N$, lifetime $\tau$, and characteristic velocity dispersion $\langle v \rangle$, calling for the construction of an analytic framework that can unify the relevant dynamical regimes.


The aforementioned studies that consider the interactions of planetary systems with passing stars have primarily been done with the aid of numerical simulations. Moreover, most of these studies have focused on the strongest form of the interactions, corresponding to the closest encounters. Such an approach is largely motivated by the characteristic length-scales of the problem: the expected distances of closest approach within typical cluster environments are on the order of 100 -- 1000\,AU \citep{proszkow}, and the orbits of interest within the solar system also span this range, extending from $30\,$AU (i.e., Neptune orbit) to $\sim500-5000\,$AU (roughly corresponding to the inner Oort cloud; \citealt{Brown2004,Sheppard2019}). Additionally, the outer edges of circumstellar disks are observed to have radii $\mathcal{L}\sim100$ AU (e.g., see the review of \citealt{williams}) and thus also fall within the confines of expected periastron distances\footnote{It is worth noting that a significant fraction of young stars reside in binary systems, with the peak of the binary distribution falling at $\sim42$ AU for solar-type stars \citep{dm}.}. 

\begin{figure*}[tbp]
\centering
\includegraphics[width=\textwidth]{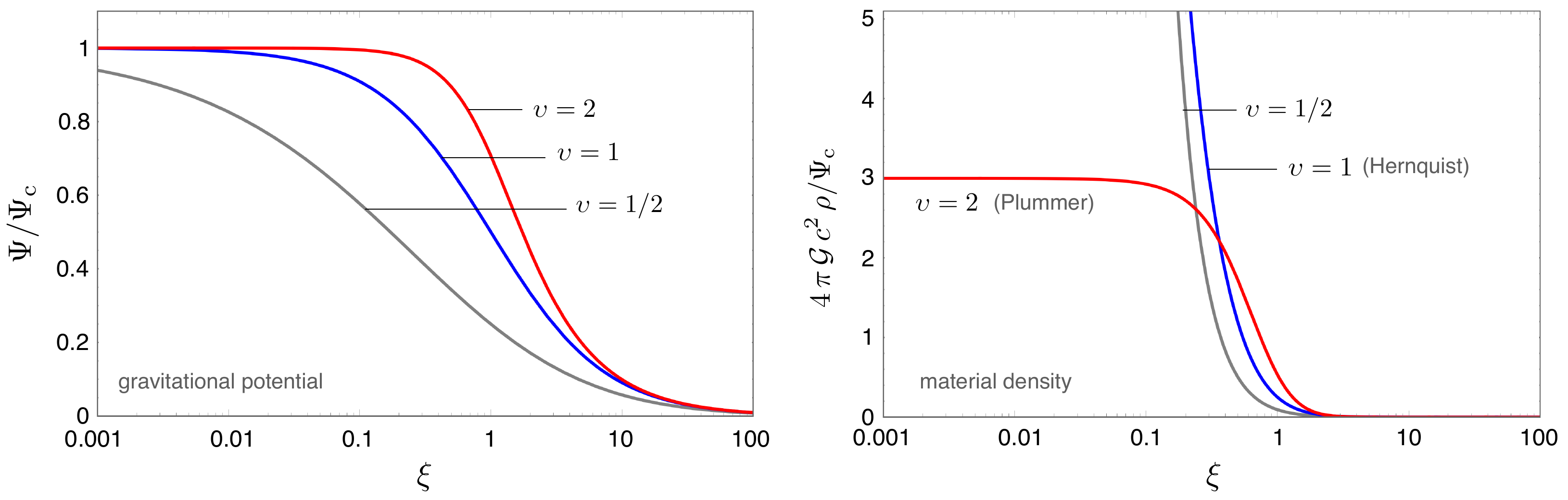}
\caption{Cluster potential--density pairs considered in this work. The left panel shows the cluster's gravitational potential (equation \ref{potential}), scaled by its central value as a function of the dimensionless radius, $\xi$. The right panel shows the corresponding scaled density profiles, which connect to the potential profiles through Poisson's equation (\ref{density}). In both panels, gray, blue, and red curves correspond to sharpness parameters of $\upsilon=1/2,\,1$ (Hernquist profile), and $2$ (Plummer profile) respectively. Note that unlike $0<\upsilon<2$ models, the $\upsilon=2$ Plummer profile yields a finite central density.}
\label{F:potential}
\end{figure*} 

The rough coincidence of these length scales (and the corresponding velocity scales) leads to hard encounters having enhanced influence \citep{al2001}. On the other hand, distant encounters are much more common, and the accumulation of their resulting weaker effects can also be important (e.g., \citealt{malmberg2011}). In this work, we develop an analytical framework to model distant encounters with passing stars as well as collective effects of the cluster, and apply our results to the trans-Neptunian region of the solar system. More specifically, we consider an orbit-averaged approach to quantifying the dynamics \citep{Rasio1995}, and limit our analysis to instances where the orbital period of the solar system objects is much shorter than the time scale of the perturbation (e.g., the time required for a fly-by encounter to take place). As we discuss below, this regime of interactions is of considerable interest for characterization of the classical Kuiper belt's evolution within the cluster. Moreover, our analytic approach allows for a greater understanding of the underlying dynamics while providing an efficient calculational framework to include the effects of many distant encounters, thus complementing numerical studies of hard (close) encounters that have been carried out previously.

For completeness, we note that in conjuction with dynamical interactions, cluster environments provide additional influences on planetary systems, including background radiation fields. In particular, massive stars within the cluster produce copious amounts of EUV and FUV radiation \citep{fatadams2008,thompson}, which can drive the evaporation of disk material (e.g., \citealt{adams2004,adams2006}). This radiation, along with X-rays that arise from more distributed sources within the cluster, also provide an important source of ionization and heating within the disk. These processes, in turn, affect disk accretion mechanisms in the early phases of evolution, and possibly even alter the chemical composition of growing planets. Although these radiative effects are important, they are beyond the scope of this present work, which focuses on gravitational dynamics. 

The remainder of this paper is structured as follows. Section \ref{section:meanfield} derives a dynamical model for the secular restricted three-body problem within a model cluster potential, and outlines a link between the ensuing dynamics and the Kozai-Lidov mechanism \citep{lidov,Kozai1962}. Section \ref{section:flybys} develops the secular approximation in the hyperbolic regime relevant to stellar flybys. Special cases are examined in section \ref{section:specialcases}, including the evolution of eccentricity enhancements of test particles, and separately, the accumulation of increases in the inclination angles. In section \ref{section:applications}, we apply this formalism to our solar system, with an emphasis on the dynamical architecture of the cold classical population of the Kuiper belt. These results place a constraint on the stellar density and lifetime of the sun's birth environment. The paper concludes in section \ref{summary} with a summary of our results and a brief discussion of their implications.

\section{Cluster Mean-Field Effects} \label{section:meanfield}

Dynamical evolution induced upon a planetary system by its host star cluster can generically be separated into two parts: mean-field effects, and stellar fly-bys. Of course, both of these classes of perturbations arise from nothing more than the gravitational potential of the stars (and, at early stages, gas) present within the cluster, but they are distinct in the length scales that they capture. Namely, mean-field effects ensue from the nearly smooth, collective potential of the distant stars within the cluster, while stellar flybys facilitate stochastic gravitational kicks from (comparatively) short-range interactions. In this section, we will focus on mean-field effects, which are simpler to quantify.

In addition to characterizing long-term evolution that results from the cluster potential, a secondary goal of this section is to delineate the relevant approximation scheme, which we will employ again in the next section, for the more involved problem of stellar flybys. Specifically, we will develop our model within a well studied framework -- the secular evolution of a test-particle, under perturbations from a distant massive body (in this case, the cluster). We note that although the original practical motivation\footnote{In a recently published paper, \citet{2019arXiv191103984I} point out that the basic structure of the Kozai-Lidov mechanism was already outlined in the work of \citet{1910AN....183..345V}.} for this now-classic problem stemmed from early spaceflight \citep{lidov}, it was quickly realized that ensuing long-term dynamics also materialize in numerous astrophysical settings, including the asteroid belt \citep{Kozai1962,Morbidelli1991}, hierarchical triple star/black-hole systems \citep{Kiseleva1998,Mardlingv2001}, and extrasolar planets \citep{Wu2003,Fab2007,Naoz2016}.

\subsection{Potential-Density Pairs} 
\label{sec:potential} 

\begin{figure*}[tbp]
\centering
\includegraphics[width=0.75\textwidth]{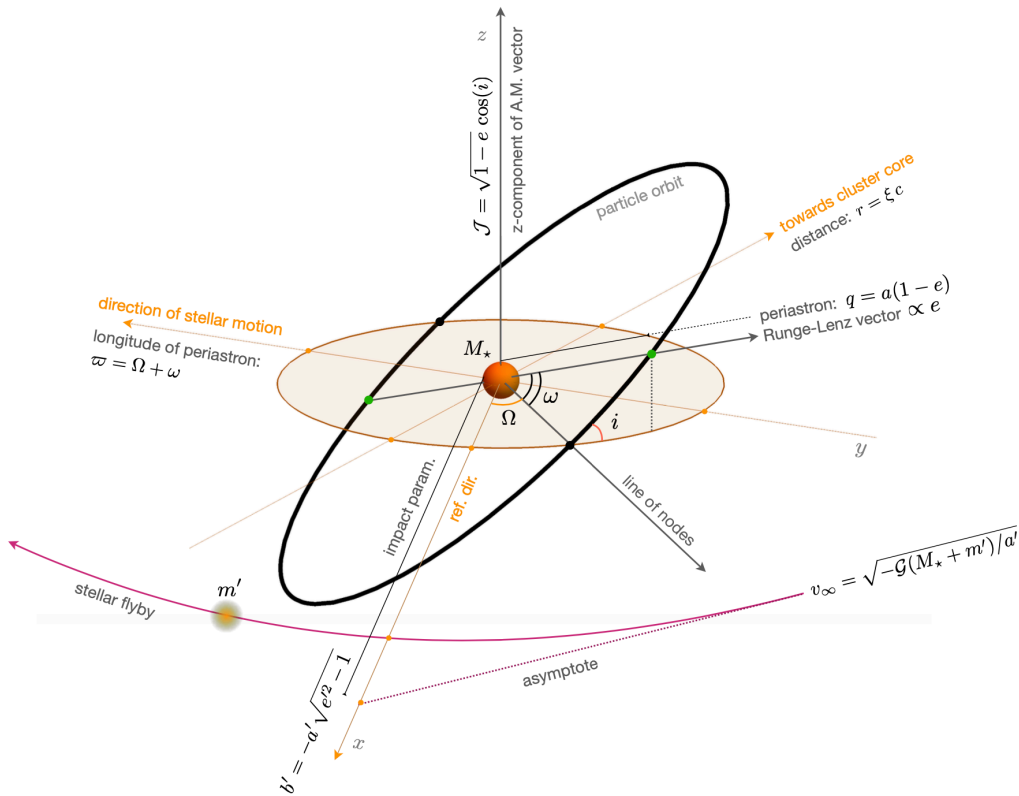}
\caption{Geometrical setup of the problem. The origin of the astro-centric coordinate system corresponds to the location of the reference star of mass $M_{\star}$. In calculations where the cluster's mean field is considered, the $z=0$ plane is taken to coincide with the orbit of the reference star within the cluster, thus defining the inclination $i$ and argument of pericenter $\omega$. As discussed in the text, the doubly phase-averaged dynamics in this case are parameterized by the normalized vertical component of the angular momentum, $\mathcal{J}$. In calculations where perturbations due to passing stars are considered, the reference plane is taken to correspond to the plane of the hyperbolic trajectory, and the reference direction is chosen to point towards the flyby's point of closest approach. Accordingly, the longitude of ascending node $\Omega$ is measured from this axis, while the orientation of the test particle orbit's major axis (in particular, the periastron) is informed by the dog-leg longitude of pericenter, $\varpi=\Omega+\omega$.}
\label{F:config} 
\end{figure*}

As a first step in quantifying long-term effects of the cluster, we must define the functional form of the cluster's gravitational potential. An archetypal model of a stellar cluster was first formulated over a century ago by \citet{Plummer1915}. Within the context of this model, the system is taken to be spherically symmetric, and the usual $\Psi\propto 1/r$ potential is softened by a characteristic length-scale, $c$, such that $\Psi$ approaches a constant value for $r\ll c$ and a point-mass potential for $r\gg c$. In the same vein, here we consider a class of softened potentials 
of the form 
\be
\Psi = - \frac{\Psi_{\rm{c}} }{ (1 + \xi^\upsilon)^{1/\upsilon} }, 
\label{potential} 
\ee
where $\xi=r/c$ is the dimensionless radius and $\Psi_{\rm{c}}\geqslant0$ by convention.

Equation (\ref{potential}) is of considerable practical interest because it corresponds to a cluster of finite mass, and simultaneously acts as a generalization of select routinely employed models from the literature. In particular, with the choice of $\upsilon=1$, we recover the Hernquist potential, and for $\upsilon=2$ we obtain the Plummer model. More generally, $\upsilon$ is a parameter that controls the sharpness of the potential turnover across the characteristic length-scale.

The radial density profile corresponding to the above potential can be easily obtained from the Poisson equation:
\be
\rho = \frac{\nabla^2\Psi}{4\,\pi\,\G}= \frac{\Psi_{\rm{c}}}{4\,\pi\,\G\,c^2}\, \frac{(1+\upsilon)}{\xi^{2-\upsilon} (1 + \xi^\upsilon)^{2+1/\upsilon}}.
\label{density} 
\ee
Figure (\ref{F:potential}) shows $\Psi$ and $\rho$ (appropriately scaled) as functions of $\xi$ for $\upsilon=1/2,1$ and $2$. It is worth noting that the $\upsilon=2$ Plummer sphere is the only model where the central density has a finite value.

Expression (\ref{density}) demonstrates that the only physically sensible choices for the sharpness parameter $c$ lie in the range $0<\upsilon\leqslant2$, since $\upsilon=0$ corresponds to constant potential (which is not of interest) and for $\upsilon>2$ the central density always approaches zero (corresponding to a Rayleigh-Taylor unstable, hollowed-out structure). At a given dimensionless radius, the enclosed mass of the cluster is determined by the integral
\be
\frac{M}{M_{\infty}} = \int_0^\xi {\xi^\upsilon d\xi \over (1 + \xi^\upsilon)^{2+1/\upsilon}} = \big(1+1/\xi^\upsilon \big)^{-(1+\upsilon)/\upsilon},
\label{mratio}
\ee
and the total mass of the system, $M_{\infty}$, is related to the potential via
\be
\Psi_{\rm{c}} = {\G\,M_\infty \over c}. 
\ee


With the relevant expressions delineated, let us now consider the characteristic quantities of a real cluster. Observational surveys indicate that the average stellar number density in clusters with $N\sim10^2-10^4$ stars is approximately $\langle \eta \rangle\sim10^2/$pc$^3$ (cluster membership-dependence of this quantity is rather weak, although radius-dependence is significant, with central values reaching upwards of $\eta_{\rm{c}}\gtrsim10^4/$pc$^3$; \citealt{hillenbrand}). As an illustrative example, we can consider a cluster with a total mass of $M_{\infty}=1200\,M_{\odot}$ (roughly comparable to the mass of the Orion Nebular Cluster) and set the mean number density of stars interior to the $M/M_{\infty}=95\%$ radius (which evaluates to $r_{95\%}=5.36\,c$ for a $\upsilon=2$ profile from equation \ref{mratio}) to $\langle \eta \rangle=100/$pc$^3$, adopting a mean IMF stellar mass of $\langle M_{\star}\rangle=0.38\,M_{\odot}$ \citep{2001MNRAS.322..231K}. This fixes the Plummer radius to $c=0.35\,$pc. In turn, this choice of parameters implies a cluster core radius of $r_{\rm{core}}=\sqrt{\sqrt{2}-1}\,c=0.23\,$pc and a central number density of $\eta_{\rm{c}}=\rho_{\rm{c}}/\langle M \rangle=1.7\times10^4/$pc$^3$. Both of these quantities are in close agreement with the properties of the Trapezium cluster (embedded within the ONC) which has a radius of $r \approx0.24\,$pc and a number density of $\eta \approx1.4\times10^4/$pc$^3$ \citep{ladalada}. 

For completeness, we note that actual clusters generally have more complicated initial conditions than those considered herein. That is, the initial states are not fully spherically symmetric, and contain substructures on a broad range of scales. As shown below, however, the effects of interest to this paper accumulate over $10-100\,$Myr, and the starting states are largely smoothed out over these timescales.

\subsection{Phase-Averaged Dynamics}

Having specified the functional form of the cluster potential in terms of physical quantities, we are now in a position to quantify the dynamical evolution induced upon a test particle orbiting a central star of mass $M_{\star}$, which itself orbits within its birth cluster at a (dimensionless) radius $\xi$. We begin by expressing the components of the astro-centric radius vector $\rr=(x,y,z)$ of the test particle in terms of Keplerian orbital elements \citep{md1999}: 
\begin{align}
&x=a \big(\cos (\mathcal{E})-e\big) \big(\cos (\omega ) \cos (\Omega )-\cos(i) \sin (\omega ) \sin (\Omega )\big)\nonumber \\
&-a \sqrt{1-e^2} \sin (\mathcal{E}) \big(\cos(i) \cos (\omega ) \sin (\Omega )+\sin (\omega ) \cos (\Omega )\big)\nonumber \\
&y=a (\cos (\mathcal{E})-e) \big(\cos(i) \sin (\omega ) \cos (\Omega )+\cos(\omega ) \sin (\Omega )\big) \nonumber \\
&+a \sqrt{1-e^2} \sin (\mathcal{E}) \big(\cos(i) \cos (\omega ) \cos (\Omega )-\sin (\omega ) \sin (\Omega )\big)\nonumber \\
&z=a \sqrt{1-e^2} \sin(i) \cos (\omega)\sin (\mathcal{E}) +a \sin(i) \sin (\omega ) \nonumber \\
&\times \big(\cos (\mathcal{E})-e\big),
\label{xyz}
\end{align}
where $a$ is the semi-major axis, $e$ is the eccentricity, $i$ is the inclination, $\omega$ is the argument of pericenter, $\Omega$ is the longitude of the ascending node, and $\mathcal{E}$ is the eccentric anomaly. For simplicity, we restrict the orbit of the central star within the cluster to the reference plane, and assume that it is circular\footnote{Lifting the assumption of a circular orbit introduces octupole-level terms into the secular Hamiltonian. Because our analysis is carried out only to quadrupolar order, the assumption of a circular orbit is not strongly limiting.} (Figure \ref{F:config}). In the frame of the central star, we then have (e.g. \citealt{ToumaWisdom1998})
\begin{align}
x'=a' \cos (\mathcal{M}') &&y'=a' \sin (\mathcal{M}') &&z'=0,
\label{xyzprime}
\end{align}
where $a'=\xi\,c$ and $\mathcal{M}'$ is the central body's mean anomaly (as measured from the cluster's center). 


Following \citet{1962AJ.....67..300K}, we define the semi-major axis ratio $\alpha=a/a'<1$ as a small parameter\footnote{An alterantive approach would be to take the ratio $a/c$ as a small parameter. The two approaches give equivalent results.} inherent to the problem, and expand $\Psi$ as a power-series in $\alpha$. The first relevant term appears at second order in $\alpha$:
\begin{align}
&\Psi^{(2)}=-\Psi_{\rm{c}}\, \alpha ^2 \left(\left(a'/c\right)^{\upsilon}+1\right)^{-(2+1/\upsilon)} \left(a'/c\right)^{\upsilon }\big(8a'^6\big)^{-1}\nonumber \\
&\times\big[4 a'^4 (\upsilon +1) \big(a'/c\big)^{\upsilon } \big(\sin  (\mathcal{M}') \big(a' \sqrt{1-e^2} \sin (\mathcal{E}) \nonumber \\
&\times(\cos(i) \cos (\omega ) \cos (\Omega ) -\sin (\omega ) \sin (\Omega )) \nonumber \\
&+a' (\cos (\mathcal{E})-e)  (\cos(i) \sin (\omega ) \cos (\Omega )+\cos (\omega ) \sin (\Omega ))\big) \nonumber \\
&+\cos (\mathcal{M}') \big(a' (\cos (\mathcal{E})-e) (\cos (\omega ) \cos (\Omega ) \nonumber \\
&-\cos(i) \sin (\omega ) \sin (\Omega ))-a' \sqrt{1-e^2} \sin (\mathcal{E}) \nonumber \\
&\times (\cos(i) \cos (\omega ) \sin (\Omega )+\sin (\omega ) \cos (\Omega ))\big)\big)^2 \nonumber \\
&+a'^2 \big(\big(\frac{a'}{c}\big)^{\upsilon }+1\big) \big(4 a'^2 (1-\upsilon ) \big(\sin (\mathcal{M}') \nonumber \\
&\times \big(a'  \sqrt{1-e^2} \sin (\mathcal{E}) (\cos(i) \cos (\omega ) \cos (\Omega )-\sin (\omega  ) \sin (\Omega )) \nonumber \\
&+a' (\cos (\mathcal{E})-e) (\cos(i) \sin (\omega ) \cos  (\Omega )+\cos (\omega ) \sin (\Omega ))\big) \nonumber \\
&+\cos (\mathcal{M}') \big(a' (\cos (\mathcal{E})-e) (\cos (\omega ) \cos (\Omega ) \nonumber \\
&-\cos(i) \sin (\omega ) \sin (\Omega )) -a' \sqrt{1-e^2} \sin (\mathcal{E})  \nonumber \\
&\times (\cos(i) \cos (\omega ) \sin (\Omega )+\sin (\omega ) \cos (\Omega ))\big)\big)^2  \nonumber \\
&-4 a'^2 \big(a'^2 \sin(i)^2 \big(\sqrt{1-e^2} \sin (\mathcal{E}) \cos (\omega )  \nonumber \\
&+\sin (\omega ) (\cos  (\mathcal{E})-e)\big)^2+\big(a' \sqrt{1-e^2} \sin (\mathcal{E}) \nonumber \\
&\times (\cos(i) \cos (\omega ) \cos (\Omega )-\sin (\omega ) \sin (\Omega )) \nonumber \\
&+a' (\cos (\mathcal{E})-e) (\cos(i) \sin (\omega ) \cos (\Omega )+\cos (\omega ) \sin (\Omega ))\big)^2 \nonumber \\
&+\big(a' \sqrt{1-e^2} \sin (\mathcal{E}) (\cos(i) \cos (\omega )  \sin (\Omega )+\sin (\omega ) \cos (\Omega ))\nonumber \\
&+a' (e-\cos (\mathcal{E})) (\cos  (\omega ) \cos (\Omega )-\cos(i) \sin (\omega ) \sin (\Omega ))\big)^2\big)\nonumber \\
&+\big(2 a' \sin (\mathcal{M}') \big(a' \sqrt{1-e^2}  \sin (\mathcal{E}) (\cos(i) \cos (\omega ) \cos (\Omega ) \nonumber \\
&-\sin (\omega ) \sin (\Omega ))+a' (\cos (\mathcal{E})-e) (\cos(i) \sin (\omega ) \cos (\Omega  ) \nonumber \\
&+\cos (\omega ) \sin (\Omega ))\big)+2 a' \cos (\mathcal{M}') \big(a' (\cos (\mathcal{E})-e) \nonumber \\
&\times (\cos (\omega ) \cos (\Omega )-\cos(i) \sin (\omega ) \sin  (\Omega ))\nonumber \\
&-a' \sqrt{1-e^2} \sin (\mathcal{E}) (\cos(i) \cos (\omega ) \sin (\Omega )\nonumber \\
&+\sin (\omega ) \cos (\Omega ))\big)\big)^2\big)\big]
\label{secondorder}
\end{align}
Although cumbersome, this expression has a well-defined physical meaning: $\Psi^{(2)}$ represents the quadrupolar component of the cluster's gravitational potential. 

Under the assumption that the test particle's motion around the star and the star's motion around the cluster core are not locked into any discernible mean-motion resonance, we may employ the secular approximation, and average $\Psi^{(2)}$ (which is the negative disturbing function) over the mean anomalies of the star and the test particle. Because the action conjugate to the mean anomaly of the test particle is solely a function of $a$, the averaging procedure results in the semi-major axis being a constant of motion. As a consequence, under this approximation, the Keplerian term of the full Hamiltonian can be dropped (e.g., \citealt{Touma2009}), implying that for the problem of interest, $\bar{\bar{\mathcal{H}}}\rightarrow\bar{\bar{\Psi}}$ (where the double over-bar signifies phase-averaging over the both the particle's and the star's orbits).

While equation (\ref{secondorder}) is expressed in terms of the test particles's eccentric anomaly, $\mathcal{E}$, the averaging procedure must be carried out in terms of the mean anomaly, $\mathcal{M}$. The two quantities are related through Kepler's equation 
\begin{align}
\mathcal{M}=\mathcal{E}-e\sin(\mathcal{E}).
\label{Keplerseqn}
\end{align}
Taking a derivative of both sides yields the Jacobian necessary to carry out the averaging process in terms of $\mathcal{E}$. With all the relevant parameters defined, we have
\begin{align}
\bar{\bar{\Ham}}&=\frac{1}{4\pi^2}\oint \oint \Psi^{(2)} \big(1-e\cos(\mathcal{E})\big)d\mathcal{E}\,d\mathcal{M}' \nonumber \\
&=\frac{\mathcal{G}\,M_{\infty}}{32\,c}\frac{\xi^{\upsilon-2} \big(a/c\big)^2}{\big(1+\xi^\upsilon \big)^{2+1/\upsilon}}\bigg[ \big(2+3\,e^3\big)\big(2+3\upsilon-\xi^\upsilon) \nonumber \\
&- \big(3\xi^\upsilon-\upsilon+2 \big)\big((2+3\,e^2)\cos(2\,i)\nonumber \\
&+10\,e^2\sin^2(i)\cos(2\omega)  \big) \bigg].
\label{Kozai}
\end{align}
Simplified expressions for $\bar{\bar{\Ham}}$ are provided in Appendix \ref{appendix:meanfield} for the specific choices of $\upsilon=1$ (Hernquist) and $\upsilon=2$ (Plummer).

The resulting Hamiltonian displays many of the same characteristics as the well-known Kozai-Lidov Hamiltonian (e.g., \citealt{KinoshitaNakai,2019MNRAS.488.5489H}). That is, Hamiltonian (\ref{Kozai}) depends on the argument of periastron, $\omega$, but not the longitude of ascending node, $\Omega$, which renders its conjugate action $\mathcal{J}=\sqrt{1-e^2}\,\cos(i)$ an integral of motion\footnote{The physical meaning of $\mathcal{J}$ corresponds to the $\hat{z}$-component of the test particle's angular momentum vector, as defined by the plane of the orbit of the central star within the cluster (see Figure \ref{F:config}).}. As a consequence, dynamical evolution facilitated by equation (\ref{Kozai}) can simply be understood by projecting level curves of $\bar{\bar{\Ham}}$ onto the $e-\omega$ plane, for a specified value of $\mathcal{J}$. In turn, by evaluating $\mathcal{J}$ at $e=0$, we can obtain a maximal value of the inclination, $i_{\rm{max}}$, attainable on a given diagram (see e.g., \citealt{Morbybook}, Ch. 8).

For the standard Kozai-Lidov resonance, the topology of the phase-space portrait is independent of the orbital separation, since this value only appears in the pre-factor of the Hamiltonian and thus only regulates the secular frequency \citep{Fab2007}. This characteristic is shared by Hamiltonian (\ref{Kozai}) in the limit of $\xi\rightarrow \infty$ (wherein the cluster is taken to be distant enough to effectively act as a faraway point-mass). In the $\xi\lesssim1$ limit on the other hand, the structure of the phase space portrait itself is determined by $\xi$, and for certain parameter combinations, the typical feature of Kozai-Lidov dynamics, where the $e=0$ equilibrium becomes secularly unstable below a critical value of $\mathcal{J}$, vanishes (see also \citealt{Brasser2006,2019MNRAS.488.5489H} and references therein). An example of this behavior can be easily demonstrated for the Plummer profile.


\begin{figure*}[tbp]
\centering
\includegraphics[width=0.75\textwidth]{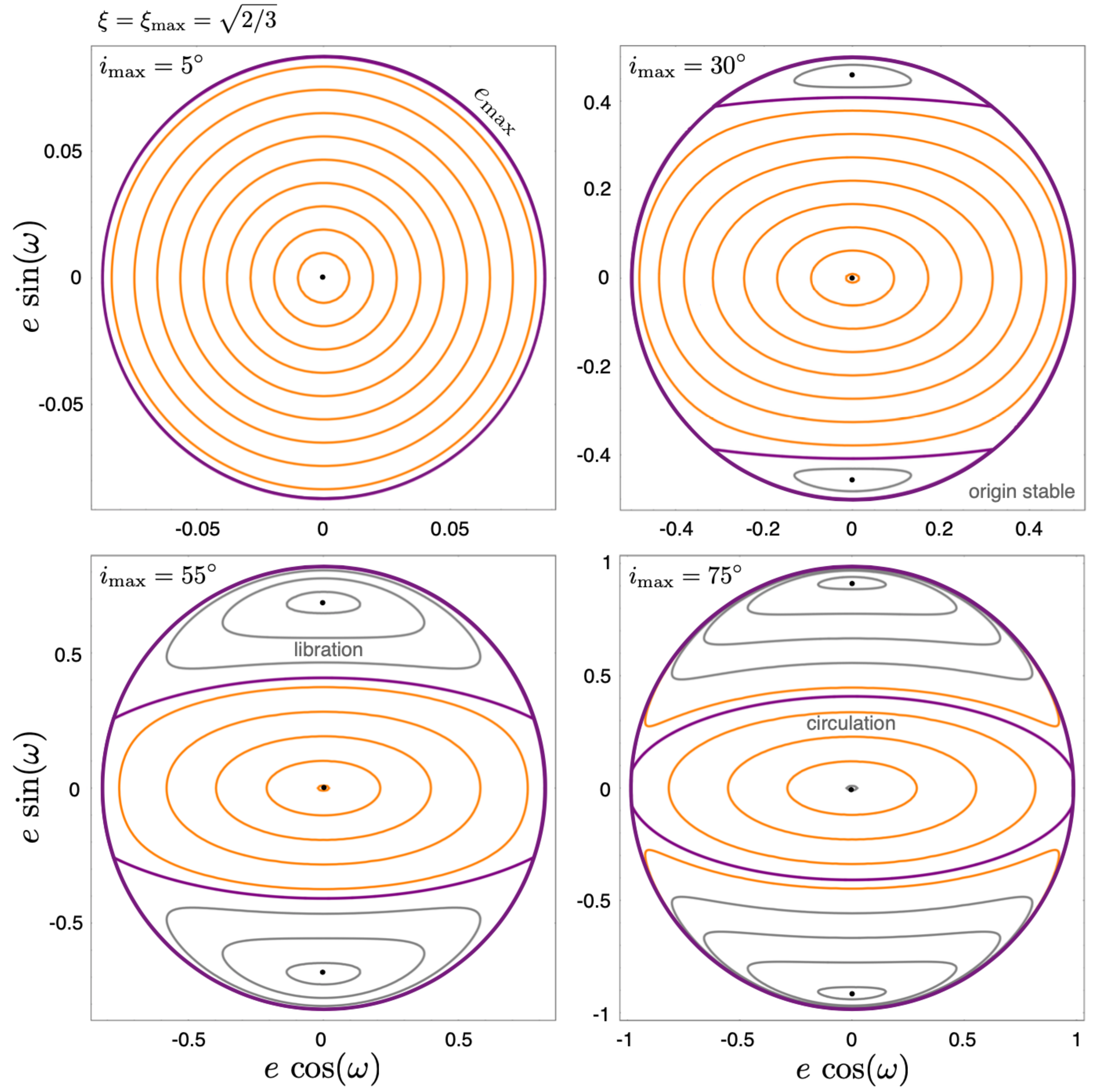}
\caption{Phase space portraits corresponding to the orbit-averaged evolution of a test-particle perturbed by the mean-field potential of the cluster. Each panel depicts the level curves of Hamiltonian (\ref{Kozai}), parameterized by a unique value of the integral of motion $\mathcal{J}=\cos(i_{\rm{max}})$. The origin of each diagram corresponds to a circular test particle orbit, while the maximal attainable eccentricity in each portrait is limited by the conservation of $\mathcal{J}$, such that $e_{\rm{max}}=\sqrt{1-\cos(i_{\rm{max}})}$. Secular trajectories corresponding to libration of $\omega$ are shown in gray, while those that exhibit $\omega-$circulation are shown in orange. In this example, the cluster is assumed to follow the $\upsilon=2$ Plummer profile, and the central star is taken to reside at a dimensionless radius $\xi=\sqrt{2/3}$, where the rate of cluster-induced secular dynamics is maximized. In contrast with the standard picture of the  Kozai-Lidov resonance, note that for this specific combination of parameters the circular orbit is secularly stable for all values of $\mathcal{J}$.}
\label{F:kozai} 
\end{figure*}

Examining equation (\ref{PlummerHam}), it is easy to see that the pre-factor of the Hamiltonian has a well-defined maximum at $\xi_{\rm{max}}=\sqrt{2/3}$, where rate of cluster-induced secular evolution is fastest. Setting $\xi=\xi_{\rm{max}}$ and $\upsilon=2$, we plot the level curves of the mean-field Hamiltonian (\ref{Kozai}) of the Plummer cluster model in Figure (\ref{F:kozai}). The four panels shown on the Figure depict the topology of $\bar{\bar{\Ham}}$ for $i_{\rm{max}}=5, 30, 55,$ and $75\deg$ in terms of the rectangular coordinates $(e\cos(\omega), e\sin(\omega))$. As is usual for Kozai-Lidov type dynamics, we see the emergence of a broad second-order secular resonance with elliptic equilibrium points located at $\omega=90\deg$ and $\omega=270\deg$ \citep{Kozai1962,Morbidelli1991}. However, unlike the standard Kozai-Lidov picture, the circular orbit does not become unstable for any value of $i_{\rm{max}}$. We emphasize that this secular stability of the circular orbit is not a generic feature of Hamiltonian (\ref{Kozai}), and is instead a consequence of the specific choice of $\xi=\xi_{\rm{max}}$ and $\upsilon=2$. Indeed, for a broad range of other parameter combinations, the $e=0$ equilibrium can be rendered hyperbolic above a critical inclination, whose value itself depends on $\xi$ (see appendix \ref{appendix:meanfield} for an illustration).

Concisely speaking, the analysis presented in this section points to the fact that the smooth component of the cluster potential can have a considerable impact on modulating the orbital eccentricities of secondary bodies, but this effect is a sensitive function of both the orbital separation of the particle from its host star as well as the location of the star within the cluster. At the same time, we note that Kozai-Lidov type dynamics is notoriously susceptible to suppression by external (e.g., planetary) sources of periapse precession, which -- if strong enough -- can trivialize the phase space portrait to resemble the $i_{\rm{max}}=5\deg$ panel of Figure (\ref{F:kozai}), for all values of $\mathcal{J}$; (e.g., \citealt{batygin2001}). This suggests that within the early solar system, the class of objects whose eccentricities could have been appreciably affected by the smooth component of the cluster potential is restricted to the long-period tail of the primordial scattered disk i.e., the Sedna population\footnote{It is worth noting that the Sedna population is thought to predate the formation of the Oort cloud, and unlike the majority of Kuiper belt objects, was likely emplaced into its current orbital neighborhood before the dissipation of the proto-solar nebula \citep{MorbyDavid2019}.} (where $a\sim500$ AU and period $P\sim10,000\,$years; \citealt{MorbidelliLevison2004,Brasser2006}). For the remainder of the solar system, the effect of the cluster was likely limited to slow rotation of the total angular momentum vector, which occurs even if the Kozai-Lidov $\omega-$resonance itself is fully suppressed. We will revisit these effects again in section \ref{section:applications}.



\section{Secular Theory of Stellar Flybys}  \label{section:flybys}

Let us now shift our focus away from the cluster's collective potential and consider the gravitational effects of passing stars. Traditionally, the motivation for understanding stellar perturbations upon planetary systems stemmed from the need to characterize cometary dynamics \citep{oort,1987AJ.....94.1330D,1988ApJ...328L..69D}. By now, there exists a rich literature on the interactions between long-period comets (and wide binaries in general) and stellar encounters (see e.g. \citealt{HeislerTremaine1986,Kaib2013,Torres2019} and the references therein). A typical approach to modeling the energy/angular momentum drift of long-period comets due to stellar encounters invokes the impulse approximation (e.g., \citealt{BT87}, Ch. 7), under the assumption that the rate of encounters is sufficiently large so that numerous encounters occur over the course of a single orbital period.

Unlike the cometary case, the effects of passing stars upon planets in young clusters lies in the regime where a single encounter occurs over numerous planetary orbital periods (in other words, the period hierarchy is switched; \citealt{Rasio1995}). In this case, the impulse approximation is not applicable, and it is sensible to instead employ the secular approximation for the planet (which we can securely treat as a test-particle) as above, and consider an averaged description of the orbital dynamics \citep{Hamers2018}. In addition to the obvious requirement that $a\ll q'=a'\,(1-e')$, a crude criterion for this approximation to hold can be written as: 
\begin{align}
&\mathcal{T}_{\rm{enc}}\sim\frac{2\,b'}{\langle v\rangle}\gg \frac{2\,\pi}{n}=P,
\label{secularcrit}
\end{align}
where $b'$ is the impact parameter of the encounter, $\langle v\rangle$ is the characteristic velocity dispersion of the cluster, and $n$ is particle's the mean motion. As an example, note that in young embedded clusters, $\langle v\rangle\sim1\,$km/s, which means that the characteristic timescale for an encounter with $b'\sim500\,$AU (approximately the semi-major axis of Sedna; \citealt{Brown2004}) is of order $\mathcal{T}_{\rm{enc}}\sim5000\,$years -- more than an order of magnitude longer than Neptune's orbital period. Obviously, more distant encounters satisfy the above criterion (\ref{secularcrit}) even better.

To a reasonable degree of accuracy, stellar flybys within a birth cluster can be assumed to be isotropically distributed. Accordingly, one avenue towards modeling the effects of individual encounters is to define an inertial coordinate system, and to follow the evolution of a particle's orbit, subject to hyperbolic perturbations arising from random directions. A physically equivalent, but more mathematically advantageous route, is to rotate the coordinate system to coincide with the orbital plane, as well as the perihelion direction of the encounter, and compute the changes in the particles' eccentricity (Runge-Lenz) as well as angular momentum vectors, assuming that the particle orbit itself is randomly oriented. This is the approach we adopt herein.

Without loss of generality, we can consider a reference frame where the $\hat{z}$-axis is orthogonal to the plane of the perturbing star's orbit, and the $\hat{x}$-axis corresponds to the direction of closest approach between the two stars (Figure \ref{F:config}). The components of the perturbing object's stellocentric radius vector are then
\begin{align}
&x'=a' \, \big( \cosh (\mathcal{W}') -e' \big) \nonumber \\
&y'=a' \, \sqrt{e'^2-1} \, \sinh (\mathcal{W}'),
\label{xyzprimehyp}
\end{align}
where $\mathcal{W}'$ is the hyperbolic eccentric anomaly\footnote{Note that unlike the elliptic eccentric anomaly $\mathcal{E}\in(0,2\pi]$, the hyperbolic eccentric anomaly $\mathcal{W}\in(-\infty,\infty).$} and as before, we set $z'=0$. 

With this definition, we follow the same procedure as in the preceeding section - namely, we expand the perturber-particle potential $\Phi=-\G\,m'/|\mathbf{r}-\mathbf{r}'|$ in powers of the ratio of characteristic length-scales. For consistency with the previous section, we retain the definition of $\alpha=a/a'$ as the small parameter inherent to the problem, but remark that developing the expansion of $\Phi$ in the ratio of particle semi-major axis to perturber impact parameter, $a/b'$, yields identical results. To this end, we further note that for $e'>\sqrt{2}$ and $e'>2$, the perturber's impact parameter and periastron distance exceed its semi-major axis, respectively. 


As in equation (\ref{secondorder}), the first relevant term in the expansion of the potential appears at second order in $\alpha$. Averaging $\Phi^{(2)}$ over the planetary mean anomaly, $\mathcal{M}$, we have:
\begin{align}
&\bar{\Ham}=\frac{1}{2\pi}\oint \Phi^{(2)}\big(1-e\cos(\mathcal{E})\big)d\mathcal{E} \nonumber \\
&= \frac{\G\,m'\,\alpha^2}{4\,a'\,(e'\cosh(\mathcal{W}')-1)^3} \Bigg[\bigg(\frac{e'-\cosh(\mathcal{W}')}{e'\cosh(\mathcal{W}')-1} \bigg)^2 \nonumber \\
&\times \bigg( 3 \left(1-e^2\right) (\cos (i) \cos (\omega ) \sin (\Omega ) \nonumber \\
&+\sin (\omega ) \cos (\Omega ))^2 +\left(12\, e^2+3\right) (\cos (\omega ) \cos (\Omega )\nonumber \\
&-\cos (i) \sin (\omega ) \sin (\Omega ))^2  \bigg) -(2+3e^2) \nonumber \\
&+\bigg(\frac{3\,\big(\cosh(\mathcal{W}'-e')\big)\sinh(\mathcal{W}')\,\sqrt{e'^2-1}}{2\big(e'\cosh(\mathcal{W}')-1 \big)^2} \bigg)\nonumber \\
&\times \bigg( \cos ^2(i) \sin (2\, \Omega ) \left(5 e^2 \cos (2 \, \omega )-3 e^2-2\right)\nonumber \\
&+10 \,e^2 \cos (i) \sin (2 \, \omega ) \cos (2 \, \Omega )+\sin (2 \, \Omega ) \big(5\, e^2 \cos (2 \, \omega )\nonumber \\
&+3\, e^2+2\big) \bigg) +\bigg(\frac{3\,\sinh^2(\mathcal{W}')\,(1-e'^2)}{2\,\big(e'\cosh(\mathcal{W}')-1\big)^2}\bigg) \nonumber \\
&\times \bigg(\cos ^2(i) \cos ^2(\Omega ) \left(5\,e^2 \cos (2 \omega )-3 \,e^2-2\right) \nonumber \\
& -5\, e^2 \cos (i) \sin (2 \omega ) \sin (2 \Omega ) \nonumber \\
&-\sin ^2(\Omega ) \left(5\, e^2 \cos (2 \omega )+3 e^2+2\right) \bigg)\Bigg].
\label{secondorderave}
\end{align}
Importantly, in addition to the secular degrees of freedom of the planetary orbit related to ($e,\omega$) and ($i,\Omega$) variable pairs, this Hamiltonian also possess implicit time dependence that enters through the hyperbolic eccentric anomaly of the passing star, $\mathcal{W}'$.

Ultimately, the primary goal of the envisioned calculation is to compute the cumulative changes in the orbital parameters of the planet due to a stellar encounter with a given geometry. In order to do this, we introduce scaled Delaunay action-angle coordinates 
\begin{align}
&G=\sqrt{1-e^2} &g=\omega \nonumber \\
&H=\sqrt{1-e^2}\cos(i) &h=\Omega.
\label{Delaunary}
\end{align}
In contrast to the standard expression for these coordinates (see e.g., \citealt{md1999}, Ch. 2; \citealt{Morbybook}, Ch. 1), the above variables have been reduced by a factor of $\sqrt{\G\,\M\,a}$. Correspondingly, in order to maintain symplecticticity, we must also divide the averaged Hamiltonian itself by the same constant factor (recall that the semi-major axis is rendered invariant by phase-averaging): $\hat{\bar{\Ham}}=\bar{\Ham}/\sqrt{\G\,\M\,a}$. 


In principle, it is possible to compute the changes in the orbital elements of the test particle by applying Hamilton's equations to $\hat{\bar{\Ham}}$ (expression \ref{secondorderave}), and integrating the resulting coupled ODEs with respect to $\mathcal{W}'$. Indeed, this approach can yield accurate results at a decreased computational cost, compared with direct numerical integration \citep{Rasio1995}. However, this procedure is cumbersome and offers little insight into the governing dynamics beyond that which can be obtained through the $N$-body route. Fortunately, for the problem at hand, we can take an additional step to further simply the Hamiltonian. In particular, we invoke a second separation of timescales, wherein the secular evolution induced upon the test particle by the stellar encounter is envisioned to operate on a much longer timescale than the flyby time itself. In other words, we assume that numerous stellar flybys are required to precess the secular angles $\omega$ and $\Omega$ by $2\pi$, such that
\begin{align}
\bigg(\frac{2\pi}{\Delta\,\omega/\mathcal{T}_{\rm{enc}}},\frac{2\pi}{\Delta\,\Omega/\mathcal{T}_{\rm{enc}}}\bigg) \gg \frac{2\,b'}{\langle v\rangle}\gg \frac{2\,\pi}{n}.
\label{timehierarchy}
\end{align}

If the timescale hierarchy (\ref{timehierarchy}) holds, then (to leading order) we can hold the particle orbit fixed over the encounter, and integrate the Hamiltonian over the encounter before deriving the equations of motion. In this way, application of Hamilton's equations to the time-integrated Hamiltonian yields a discrete mapping that transforms the unperturbed test-particle orbit to its post-encounter state \citep{LL1983}. Accordingly, we arrive at the cumulative changes in the Delaunary actions in the following manner:
\begin{align}
&\Delta G=-\int_{-\infty}^{\infty} \frac{\partial\hat{\bar{\Ham}}}{\partial g}\,dt \rightarrow -\frac{\partial}{\partial \omega} \int_{-\infty}^{\infty}\hat{\bar{\Ham}}\,dt = - \frac{\partial\bar{\bar{\Kam}}}{\partial \omega} \nonumber \\
&\Delta H=-\int_{-\infty}^{\infty} \frac{\partial\hat{\bar{\Ham}}}{\partial h}\,dt \rightarrow -\frac{\partial}{\partial \Omega} \int_{-\infty}^{\infty} \hat{\bar{\Ham}}\,dt = - \frac{\partial\bar{\bar{\Kam}}}{\partial \Omega},
\label{EOM}
\end{align}
with similar expressions for the changes in the angles, $\Delta\omega$ and $\Delta\Omega$. We remark that because $\bar{\mathcal{H}}$ is a measure of orbit-averaged specific energy and $\sqrt{\G\,\M\,a}$ corresponds to the maximal specific angular momentum attainable by the test particle orbit, the reduced Hamiltonian $\hat{\bar{\mathcal{H}}}$ is a measure of secular frequency. Therefore, the time-integrated Hamiltonian $\bar{\bar{\mathcal{K}}}$ is dimensionless. 

To evaluate the integral that transforms $\hat{\bar{\mathcal{H}}}\rightarrow\bar{\bar{\mathcal{K}}}$, we employ the hyperbolic variant of Kepler's equation
\begin{align}
\mathcal{Q}'=e'\sinh(\mathcal{W}')-\mathcal{W}',
\label{hypKep}
\end{align}
where $\mathcal{Q}'=\sqrt{-\G(\M+m')/a'^3}\,t=n'\,t$ is the hyperbolic mean anomaly and $n'$ is the correspondent mean motion. This allows us to carry out the integration with respect to the hyperbolic eccentric anomaly, $d\mathcal{W}'$, with the appropriate Jacobian. The time-integrated Hamiltonian thus takes the form:
\begin{align}
&\bar{\bar{\Kam}}=\int_{-\infty}^{\infty} \hat{\bar{\mathcal{H}}} \,dt = \frac{1}{n'}\int_{-\infty}^{\infty} \bar{\Kam}\,\big(e'\cosh(\mathcal{W}'-1)\big) d\mathcal{W}' \nonumber \\
&=\frac{a^3}{16\, e'^2\,b'^3}\frac{n}{n'}\frac{m'}{M} \bigg[ \left(3 \, e^2+2\right) e'^2\, \kappa \, (3 \cos (2 i)+1) \nonumber \\
&+ 30\, e^2 \, e'^2 \, \kappa \, \sin^2(i) \cos (2 \,\omega )\nonumber \\
&+2 \left(3 \,e^2+2\right) \left(e'^2-1\right)^{3/2} \sin^2(i) \cos (2 \,\Omega ) \nonumber \\ 
&+ 5 e^2 \left(e'^2-1\right)^{3/2} (\cos (i)+1)^2 \cos (2 (\omega +\Omega )) \nonumber \\
&+5 e^2 \left(e'^2-1\right)^{3/2} (\cos (i)-1)^2 \cos (2 (\omega -\Omega )) \bigg],
\label{secularfinal}
\end{align}
where
\begin{align}
&\kappa=2\bigg[\frac{\sqrt{e'^2-1}}{2}+ \arctan\left(\frac{1}{\sqrt{e'^2-1}}\right) \nonumber \\
&+\arctan\left(\frac{e'-1}{\sqrt{e'^2-1}}\right)\bigg]\approx e'+\frac{\pi}{2}+\frac{1}{2\,e'}.
\label{kappa}
\end{align}


 
The secular harmonics of the above Hamiltonian have well-defined physical interpretations. Qualitatively, the second line of equation (\ref{secularfinal}) governs the hyperbolic variant of the Kozai-Lidov resonance discussed in the previous section. On the other hand, the term on the third line regulates the interactions between the orbital planes (equivalently angular momentum vectors) of the planet and the perturber. Finally, the last two lines of $\bar{\bar{\Kam}}$ respectively facilitate prograde and retrograde eccentricity coupling (i.e., interactions between the Runge-Lenz vectors) between the particle and the passing star. 

Physical meanings of the harmonics aside, recall that by virtue of adopting a coordinate system that is aligned with the hyperbolic orbit of the perturber, in practice, each individual encounter must be modeled assuming a new, isotropically distributed orientation of the particle orbit, which translates to correspondent random values of its inclination, argument of perihelion, and longitude of ascending node. It is further important to note that at first glance, all critical arguments other than the Kozai-Lidov angle, $2\,\omega$, in Hamiltonian (\ref{secularfinal}) appear to not satisfy D'Almbert rules. This issue is, however, illusory, and stems from our choice of coordinate system. That is, an implicit assumption of equations (\ref{xyzprimehyp}) is that both $\omega'=0$ and $\Omega'=0$, meaning that even though the harmonics $2(\Omega-\Omega')$, $2(\omega+\Omega-\omega'-\Omega')$, $2(\omega-\Omega-\omega'+\Omega')$ constitute differences of longitudes that satisfy D'Almbert rules, the primed quantities do not explicitly appear in expression (\ref{secularfinal}).

\section{Special Cases} \label{section:specialcases}

The secular flyby Hamiltonian obtained in the previous section possesses two coupled degrees of freedom, and is therefore generally not integrable \citep{Morbybook}. Nevertheless, integrability of $\bar{\bar{\Kam}}$ is still attainable under certain restrictive assumptions, and in this section we consider such simplified special cases. Although primarily of academic interest (see also \citealt{1982SvA....26..721S}), this analysis allows for an illuminating exploration of the qualitative features the emergent dynamics, and for a simple comparison between analytic and numerical results. We begin by considering a 2D configuration where the plane of the particle orbit is taken to coincide with that of the passing star's trajectory.

\subsection{Eccentricity Evolution in the Plane}

\begin{figure}[tbp]
\centering
\includegraphics[width=\columnwidth]{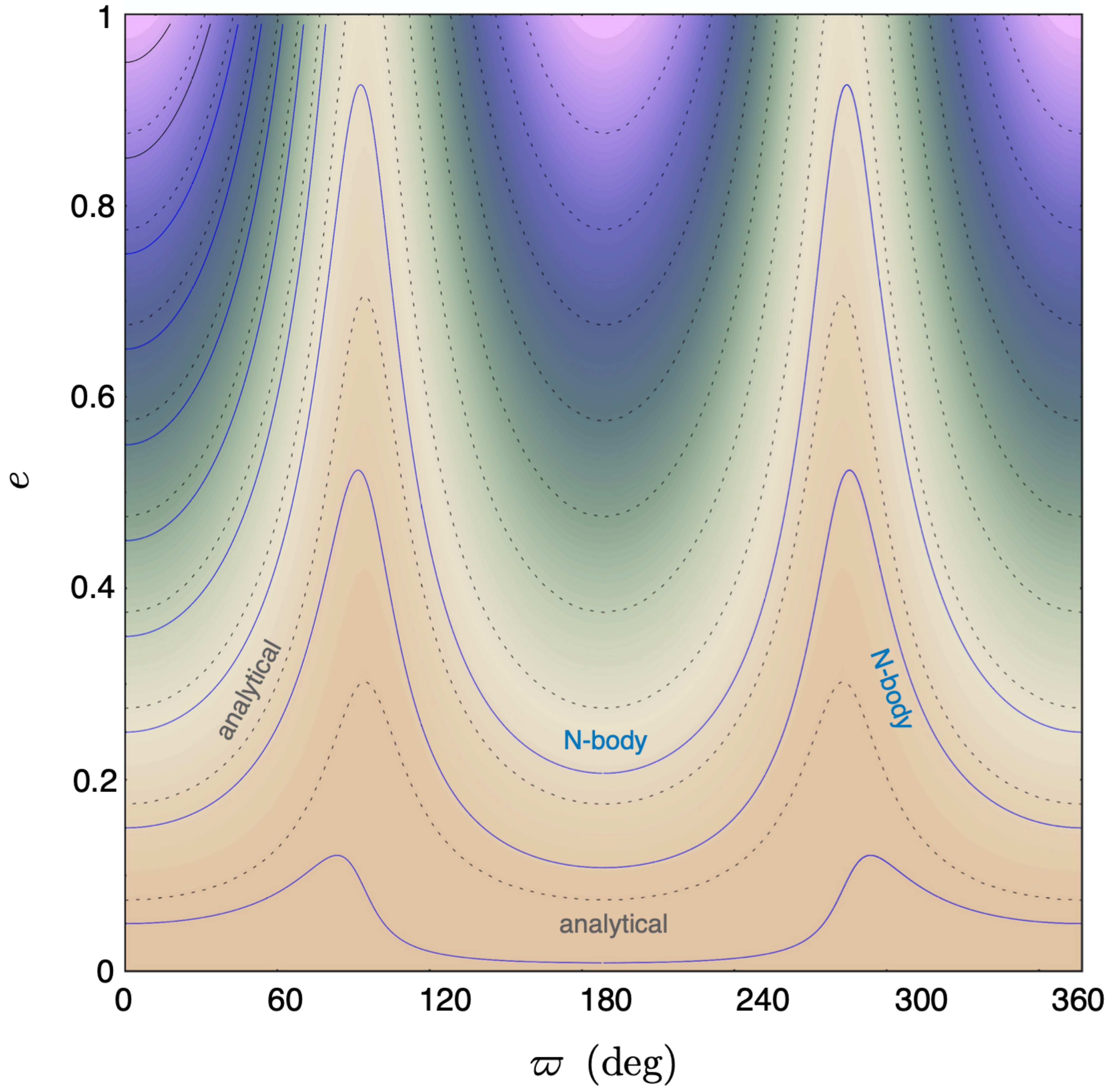}
\caption{Integrable secular dynamics corresponding to planar (2D) encounters. The figure depicts a projection of the level curves of Hamiltonian (\ref{Hamplan}) onto the $(e,\varpi)$ plane, for $e'=3$. Dashed curves as well as the background color-scale are obtained analytically, while the solid purple curves represent evolution resulting from direct $N$-body simulations of repeated encounters with $a/b'=0.035$.}
\label{F:e_integ} 
\end{figure}

Setting $i=0$ or $i=\pi$, and dropping constant terms, the Hamiltonian takes on the following rudimentary form:
\begin{align}
\bar{\bar{\Kam}}=\frac{a^3}{b'^{3}}\frac{n}{n'}\frac{m'}{M}\bigg[ \frac{3}{4} e^2 \kappa + \frac{5}{4} \frac{e^2}{e'^2}(e'^2-1)^{3/2} \cos (2\varpi) \bigg],
\label{Hamplan}
\end{align}
where $\varpi=\omega\pm\Omega$ is the longitude (as opposed to argument) of perihelion. Because the action conjugate to the angle $\gamma=-\varpi$ is the second \Poincare\ momentum $\Gamma=1-\sqrt{1-e^2}$ -- which is a sole function of $e$ -- this Hamiltonian is integrable. This means that the dynamics encapsulated by equation (\ref{Hamplan}) can be explored simply by projecting its contours onto the $e-\varpi$ plane. An illustrative example of such a projection for perturbations characterized by $e'=3$ is shown on Figure (\ref{F:e_integ}), where analytic level curves of $\bar{\bar{\Kam}}$ are depicted with dotted lines as well as the background color-scale. 

\paragraph{Comparison with $N-$body Simulations} Contours shown in Figure (\ref{F:e_integ}) provide a simple testing ground for the evaluation of assumptions inherent to the analytical model described above. In particular, our perturbative analysis suggests that a test-particle orbit subjected to repeated co-planar encounters with $e'=3$ will evolve along a secular trajectory that will trace the contours of the Hamiltonian (\ref{Hamplan}). In an effort to test this expectation, we conducted a sequence of numerical $N-$body experiments, where a test particle with initial $\varpi_0=0$ and $e_0=0.05,0.15,0.25,...,0.95$ was subjected to recurrent encounters with a $m'=\M$ perturber that followed a hyperbolic trajectory characterized by $a/b'=0.035$. The encounters were simulated such that the perturbing object would originate with a hyperbolic mean anomaly of $\mathcal{Q}'=-10^5\,$radians and persist until $\mathcal{Q}'=10^5\,$radians, after which the phase of the passing star would be abruptly re-set to its initial value, and the encounter would repeat, perturbing the orbit of the test-particle further. 

To carry out the $N$-body simulations, we used the well-tested \texttt{mercury6} gravitational dynamics software package \citep{1999MNRAS.304..793C}. The integrations were performed using the conservative variant of the Bulirsch-Stoer algorithm \citep{1992nrca.book.....P}, with an accuracy parameter set to one part in ten billion and an initial time-step equal to $0.5\%$ of the test particle's orbital period. The results from this set of numerical experiments are shown as purple curves in Figure (\ref{F:e_integ}). Clearly, the agreement between analytical and numerical results is satisfactory, although not exact: while analytical expression (\ref{Hamplan}) is exactly symmetric about $\varpi=\pi/2$, numerical results show a subtle asymmetry at low-eccentricities. It is likely that this detail can be attributed to the fact that $\bar{\bar{\Kam}}$ is a second-order Legendre polynomial expansion of the full Hamiltonian, and accounting for higher-order terms \citep{2019MNRAS.tmp.1557H} may resolve this minor discrepancy. More importantly, the confluence of analytic and numerical results depicted in Figure (\ref{F:e_integ}) illuminates an intriguing aspect of scattering dynamics -- the elliptic stability of nearly-circular obits, and an existence of a critical contour of $\bar{\bar{\Kam}}$ that divides bound and unbound evolution. Let us explore this attribute of Hamiltonian (\ref{secularfinal}) further.

\begin{figure*}[tbp]
\centering
\includegraphics[width=0.75\textwidth]{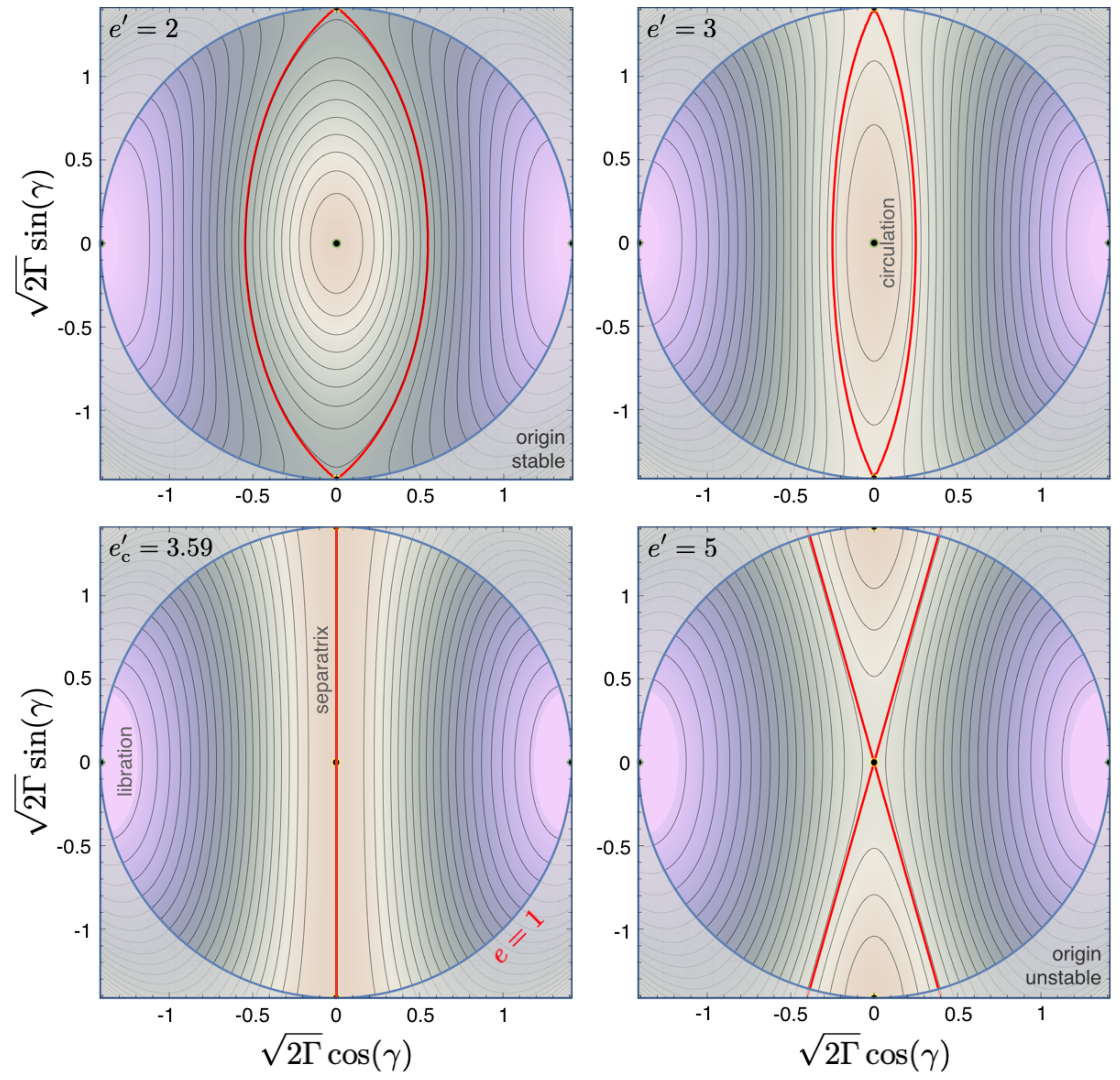}
\caption{Phase-space portraits of planar encounter dynamics in the secular regime. The level curves of Hamiltonian (\ref{Hamplanxy}) are shown in terms of cartesian analogues of the \Poincare\ action-angle coordinates, where $\Gamma=1-\sqrt{1-e^2}$ is the scaled angular momentum deficit in the plane and $\gamma=-\varpi$ is the negative longitude of pericenter. In each panel, the separatrix is shown as a bold red curve. The topology of the phase-space diagram -- and in particular the secular stability of the $e=0$ orbit (origin) -- depends on $e'$: below a critical perturber eccentricity $e'_{\rm{c}}\approx3.59$, the circular orbit corresponds to an elliptic equilibrium point in phase-space, while above the critical eccentricity, this fixed point becomes hyperbolic.}
\label{F:phase_space} 
\end{figure*} 

An interesting feature of Figure (\ref{F:e_integ}) is that only high-eccentricity elliptic orbits connect smoothly to parabolic ones. This is evident by inspection of numerical results pertaining to orbits with $e_0\geqslant 0.35$, all of which get driven upwards in $e$ as $\varpi$ precesses away from zero. On the other hand, examination of the three low-eccentricity numerical solutions shown in Figure (\ref{F:e_integ}) demonstrate that after a large number of gravitational scattering events, these orbits not only remain bound to their host star, they predictably return to their initial states. Put simply, this means that already eccentric orbits are readily made more eccentric by close encounters, while circular orbits have a tendency to remain circular. Curiously, this type of evolution signals a sharp contrast between the fundamental nature of perturbations facilitated by secular and short-periodic gravitational encounters. Specifically, while the former can lead to closed orbits in phase-space as shown in Figure (\ref{F:e_integ}), the impulsive evolution driven by the latter class of events leads to an essentially diffusive random walk through phase space, which always results in ejection, given sufficient time \citep{frozen}.

\paragraph{Secular Stability of Circular Orbits} Is elliptic stability of (nearly-)circular orbits globally ensured for all phase-averaged planar perturbations? To answer this question, let us examine the stationary solutions to Hamilton's equations in greater detail. For convenience, we appeal to canonical cartesian analogues of \Poincare\ action-angle variables (not to be confused with cartesian coordinates used in equation \ref{xyz}; \citealt{Morbybook}):
\begin{align}
x=\sqrt{2\Gamma}\cos{\gamma} &&y=\sqrt{2\Gamma}\sin{\gamma}.
\label{vars}
\end{align}
In terms of these variables, Hamiltonian (\ref{Hamplan}) reads\footnote{Interestingly, \citet{BubPetrovich2019} find an identical Hamiltonian for the planar evolution of a binary in a triaxial potential (see their equation 33).}
\begin{align}
&\bar{\bar{\Kam}}=-\frac{\alpha^3}{4\, e'^2\,(e'^2-1)^{3/2}}\frac{n}{n'}\frac{m'}{M}\bigg[ \bigg( \frac{4-x^2-y^2}{4} \bigg)\nonumber \\
&\times \big(3 \, e'^2\, \kappa \, (x^2+y^2) + 5 \, \left(e'^2-1\right)^{3/2} (x^2-y^2)\big) \bigg],
\label{Hamplanxy}
\end{align}
and its equilibria are specified by the relations
\begin{align}
\frac{d x}{d t}=-\frac{\partial \bar{\bar{\Kam}}}{\partial y} =0 && \frac{d y}{d t}=\frac{\partial \bar{\bar{\Kam}}}{\partial x} =0.
\label{equilibria}
\end{align}

In general, equations (\ref{equilibria}) admit nine solutions, but only five of them are physical. That is, Hamiltonian (\ref{Hamplanxy}) has real fixed points at $(x,y)=(\sqrt{2},0)$, $(0,\sqrt{2})$, $(-\sqrt{2},0)$, $(0,-\sqrt{2})$, and $(0,0)$. As is evident from the definitions of the variables (\ref{vars}), the equilibrium point located at the origin corresponds to a circular orbit, while the other four fixed points translate to parabolic ($e=1$) trajectories. The remaining four solutions to equations (\ref{equilibria}) all lie outside of the $x^2+y^2\leqslant2$ domain and therefore entail imaginary eccentricities. 

The Hessian matrix of $\bar{\bar{\Kam}}$, evaluated at $(x,y)=(0,0)$ reads:
\begin{align}
\mathscr{H}=\mathcal{C}
 \begin{bmatrix}
      \frac{3\,\kappa\,e'^2+5(e'^2-1)^{3/2}}{2\,e'\,(e'^2-1)^{3/2}} && 0 \\ 0 &&  \frac{3\,\kappa\,e'^2-5(e'^2-1)^{3/2}}{2\,e'\,(e'^2-1)^{3/2}}
    \end{bmatrix},
\label{Hessian}
\end{align}
where $\mathcal{C}=(a/a')^3(n/n')(m'/M)$. While the first (top left; $\partial^2\bar{\bar{\Kam}}/\partial x^2$) element of $\mathscr{H}$ is positive definite, the fourth (bottom right; $\partial^2\bar{\bar{\Kam}}/\partial y^2$) element is positive for $e'\sim1$, but negative for $e'\gg1$. This means that the secular fixed point of $\bar{\bar{\Kam}}$ that corresponds to $e=0$ is a local maximum for low $e'$, but becomes a saddle point at sufficiently large values of the perturber's eccentricity. Thus, the critical value of $e'$ at which the origin becomes a hyperbolic equilibrium is simply given by the solution to $3\,\kappa\,e'^2-5(e'^2-1)^{3/2}=0$ and quantitatively evaluates to $e'_{\rm{c}}\approx3.59$. Note that the critical value of the perturber's eccentricity does not depend on its mass, mean motion, or impact parameter, since all of these quantities appear outside of the square brackets of Hamiltonian (\ref{Hamplan}), and therefore only determine the rate at which secular evolution unfolds.

\begin{figure}[tbp]
\includegraphics[width=\columnwidth]{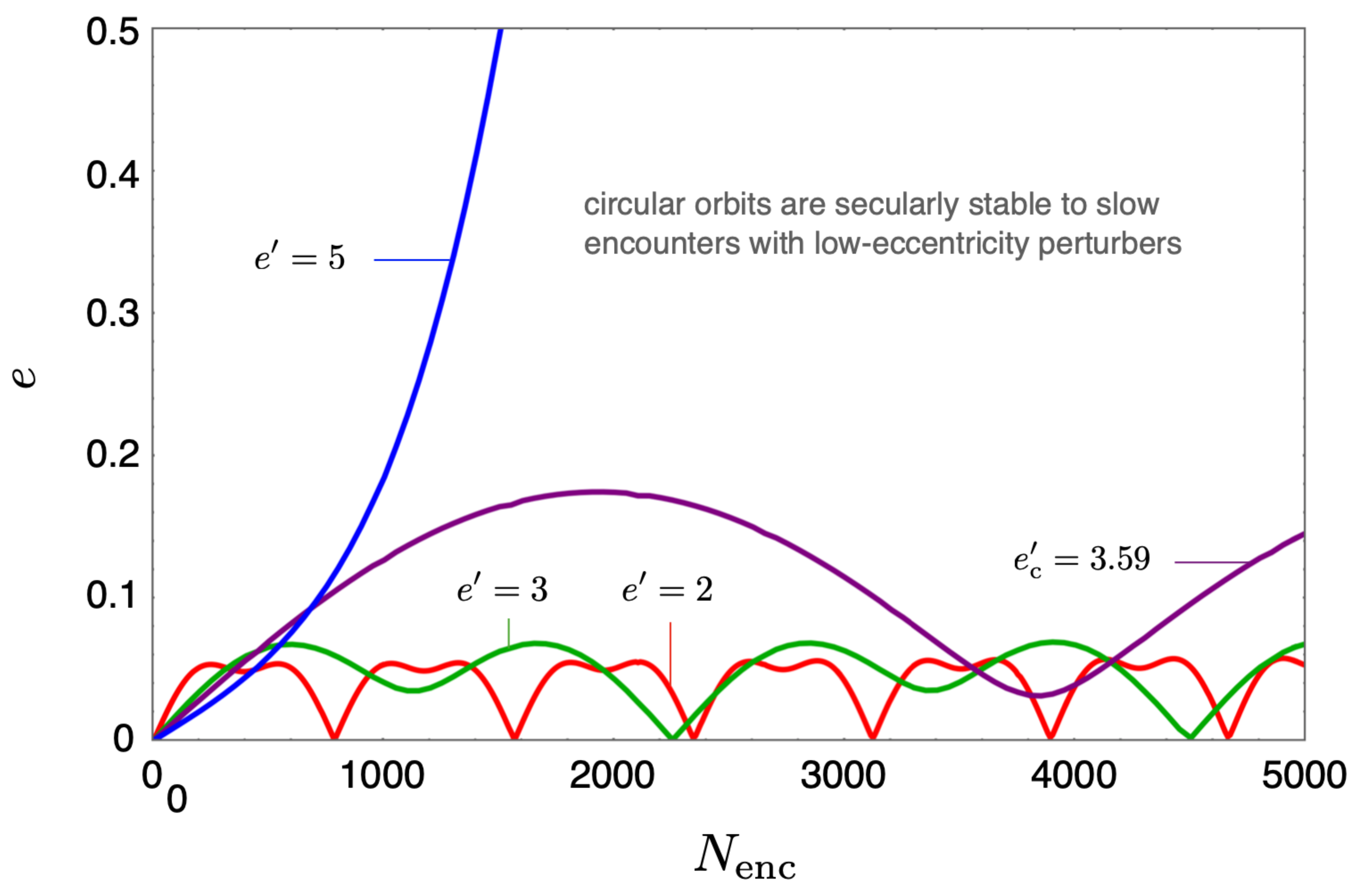}
\caption{Effective time-series of test particle evolution under repeated encounters with planar $m'=M_{\star}$ perturbers with eccentricity $e'=2$ (red), $e'=3$ (green), $e'_{\rm{c}}\approx3.59$ (purple), and $e'=5$ (blue). For all simulations, the ratio of particle semi-major axis to perturber impact parameter was set to $a/b'=0.03$. As predicted by analytic theory, when subjected to repeated perturbations from flybys with $e'\lesssim3.6$, orbits that originate with low eccentricity remain roughly circular. Conversely, for $e'\gtrsim3.6$, initially circular orbits can be rendered parabolic given a sufficient number of encounters, as demonstrated by the approximately exponential rise in eccentricity of the $e'=5$ numerical experiment.}
\label{F:e_enc} 
\end{figure} 

Figure (\ref{F:phase_space}) shows the phase-space portraits of Hamiltonian (\ref{Hamplanxy}) for a sequence of perturber eccentricities. Specifically, the four panels depict sub-critical $e'=2$ (top left panel), nearly critical $e'=3$ (top right panel), critical $e'=e'_{\rm{c}}\approx3.59$ (bottom left panel), and super-critical $e'=5$ (bottom right panel) phase-space diagrams of the test-particle. Notably, equivalent portraits with $e'$ significantly in excess of $e'_{\rm{c}}$ are qualitatively similar to the bottom right panel of Figure (\ref{F:phase_space}) and we omit them to curtail redundancy. 

To further exemplify the dependence of the $(x,y)=(0,0)$ fixed point on $e'$, we performed an additional set numerical experiments. In particular, Figure (\ref{F:e_enc}) depicts the temporal evolution of initially circular orbits, subjected to repeated encounters with $m'=\M$, $a/b'=0.03$ stars, for the same values of $e'$ as those quoted in Figure (\ref{F:phase_space}). We reiterate that the resulting evolution shown in Figure (\ref{F:e_enc}) was computed in a self-consistent $N$-body fashion as described above, rather than with the aid of our secular model. In agreement with analytic expectations, for $e'\lesssim3.6$, initially circular orbits remain nearly circular for all time, while in the simulation with $e'=5$, the circular orbit is rendered long-term unstable, achieving a parabolic shape after $N_{\rm{enc}}\approx2000$ stellar passages.

\paragraph{Critical Impact Parameter} In light of the approximation scheme employed above, it is obvious that our analytic results can only hold true as long as a leading-order expansion of the Hamiltonian in the semi-major axis ratio provides an adequate representation of the dynamics. Accordingly, before leaving this subsection, let us employ the $i=0,\pi$ special case to perform one more test, in order to determine the characteristic value of $a/b'$ at which the discrepancy between numerical and analytical results becomes large. To quantify the approximate value of $a/b'$ above which our secular formalism breaks down, we carried out a sequence of Monte Carlo simulations, comparing analytical and numerical results across a broad range of system parameters.

For definitiveness, we performed three suites of analytical and numerical simulations setting the perturber's eccentricity to $e'=2,3$, and $5$ as in Figures (\ref{F:phase_space}) and (\ref{F:e_enc}). Then, for each choice of $e'$, we simulated 2500 encounters, randomly selecting the particle's eccentricity and longitude of perihelion from uniform distributions spanning the range $e\in(0,1^{-})$; $\varpi\in(0,2\pi)$, and drawing the semi-major axis from a log-flat distribution, such that $\log_{10}a/b'\in(-2,0)$. Employing canonical cartesian analogues of equations (\ref{EOM}), we computed the analytic estimates of the changes in the canonical eccentricity vector ($\Delta x,\Delta y$)$_{\rm{an}}$ and compared them with the corresponding values computed using the direct $N$-body approach ($\Delta x,\Delta y$)$_{\rm{num}}$. We then computed the fractional error
\begin{align}
\zeta=\sqrt{ \frac{ (\Delta x_{\rm{num}}- \Delta x_{\rm{an}})^2 + (\Delta y_{\rm{num}}- \Delta y_{\rm{an}})^2  }{{\Delta x^2_{\rm{num}}+\Delta y^2_{\rm{num}}}} }
\label{error}
\end{align}
for each encounter. 

\begin{figure}[tbp]
\includegraphics[width=\columnwidth]{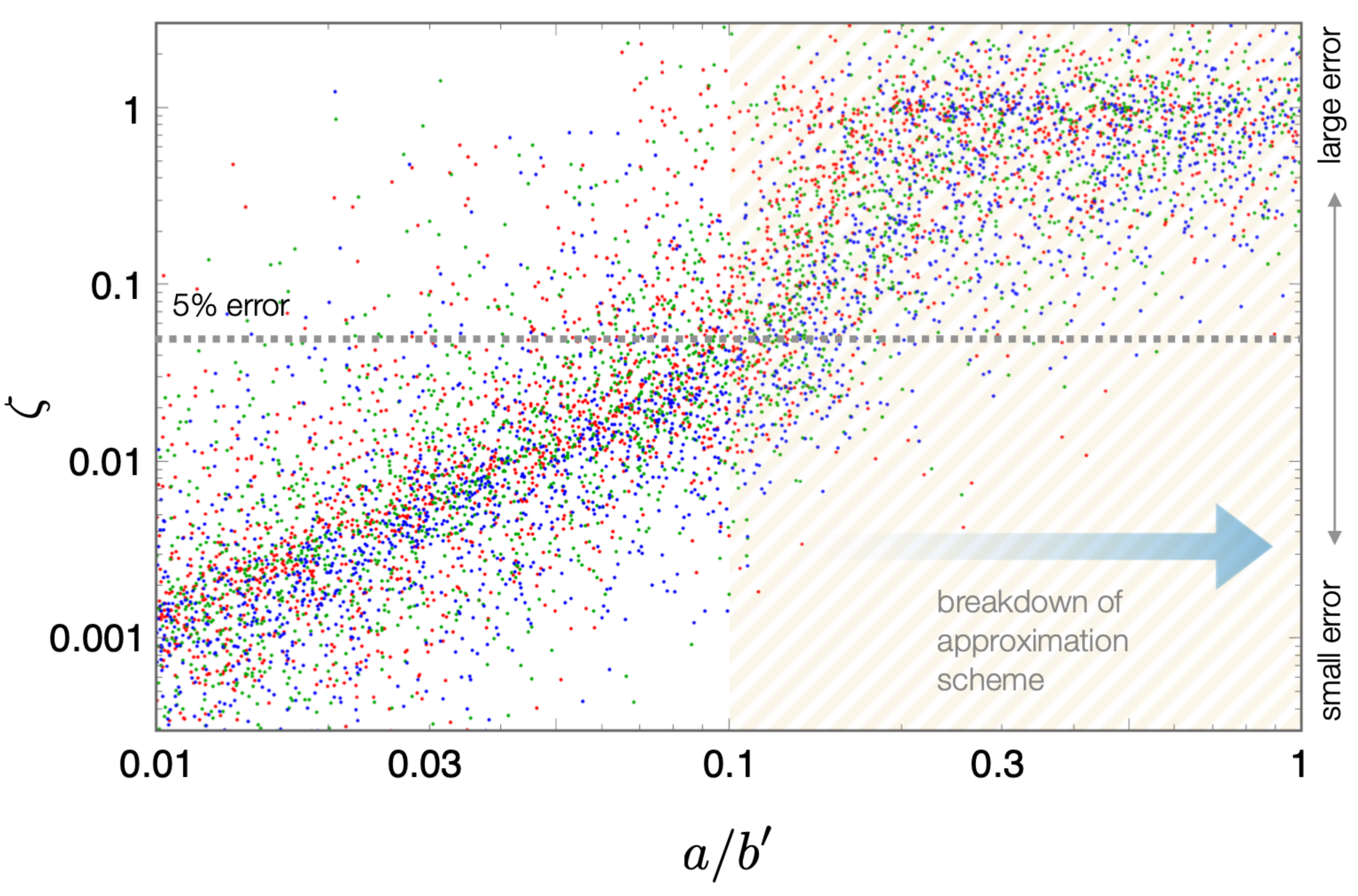}
\caption{Fractional error of the analytic approximation scheme, $\zeta$, as a function of the semi-major axis to impact parameter ratio. The figure reports the results of three sets of $N$-body simulations, with red, green, and blue points corresponding to perturber eccentricities of $e'=2$, $e'=3$, and $e'=5$ respectively. Clearly, our analytic approximation scheme becomes inadequate for semi-major axis to impact parameter ratio of $a/b'\gtrsim0.1$.}
\label{F:error} 
\end{figure} 

Figure (\ref{F:error}) shows $\zeta$ as a function of $a/b'$, where we have employed the same color scheme for perturber eccentricities as that in Figure (\ref{F:e_enc}). Overall, irrespective of $e'$, the results portray a consistent picture: the error inherent to our analytic approximation scheme is essentially negligible for $a/b'\sim0.01$ but grows approximately as $\zeta\propto(a/b')^{3/2}$, such that at $a/b'\sim0.1$, it can be as large as a few percent. Cumulatively, this analysis suggests that the secular perturbation theory employed in the derivation of Hamiltonian (\ref{secularfinal}) is adequate for impact parameters that obey $a/b'\lesssim0.1$. Given that the semi-major axes of classical Kuiper belt objects do not extend beyond $a\sim50\,$AU, $b'\sim500\,$AU represents a critical impact parameter below which application of the developed framework to the solar system becomes suspect. Notably, the minimum expected impact parameter corresponding $\eta\sim \langle \eta \rangle\approx100/$pc$^3$ and $\tau\sim100\,$Myr exceeds $b'_{\rm{min}}\gtrsim1,000\,$AU. 

\subsection{Inclination Evolution of Circular Orbits}\label{sec:inc_circ_orb}

Having just characterized coplanar encounters with eccentric perturbers, let us now consider the opposite extreme: inclined encounters with test particles on circular orbits. One astrophysically relevant setting where such dynamics emerges naturally is the evolution of protoplanetary disks residing within stellar associations. Owing to hydrodynamic forces and viscosity, fluid astrophysical neulae have a natural tendency to relax towards nearly-axisymmetric structures, justifying the $e\rightarrow0$ assumption \citep{FragnerNelson2010,2014MNRAS.440.1179X,picogna}. For definitiveness, we will begin our discussion with the simple example of a test-particle as above, and subsequently generalize our results to radially extended structures.

Setting $e=0$ and dropping constant terms, Hamiltonian (\ref{secularfinal}) simplifies to the following integrable form:
\begin{align}
&\bar{\bar{\Kam}}=\frac{a^3}{8\, e'^2\,b'^{3}}\frac{n}{n'}\frac{m'}{M}\bigg[ 3\,e'^2\,\kappa\,\cos(2i) \nonumber \\
&+2 \big(e'^2-1 \big)^{3/2} \sin^2(i)\,\cos(2\,\Omega) \bigg] \nonumber \\
&= -\frac{\alpha^3}{8\, e'^2\,(e'^2-1)^{3/2}}\frac{n}{n'}\frac{m'}{M}\bigg[3\,e'^2\,\kappa\,\big(2\,H^2-1 \big) \nonumber \\
&+2 \big(e'^2-1 \big)^{3/2} \big(1-H^2 \big)\,\cos(2\,h) \bigg].
\label{Hplanes}
\end{align}
An intriguing feature of this Hamiltonian is that for $e'\gg1$, the dependence of $\bar{\bar{\Kam}}$ on $e'$ simplifies considerably. In particular, recalling the series expansion for $\kappa$ from equation (\ref{kappa}), we have
\begin{align}
&\bar{\bar{\Kam}}=-\frac{\alpha^3}{4\, e'^2}\frac{n}{n'}\frac{m'}{M}\bigg[ 3\,H^2 +\big(1-H^2 \big)\,\cos(2\,h) \bigg].
\label{Hplanesegg1}
\end{align}

Compared with the planar special case described in the previous section, the fixed points of Hamiltonian (\ref{Hplanes}) are also considerably simpler. Specifically, noting the quadratic and cosinusoidal dependence of the Hamiltonian on $H$ and $h$ respectively, the equilibrium equations
\begin{align}
&\frac{dh}{dt}=\frac{\partial \bar{\bar{\Kam}}}{\partial H}\propto H = 0 &\frac{dH}{dt}=-\frac{\partial \bar{\bar{\Kam}}}{\partial h}\propto \sin(2\,h) = 0
\label{Hplanesegg1eq}
\end{align}
imply that all fixed points of equation (\ref{Hplanes}) reside at $i=\pi/2$ and $\Omega=0,\pi/2,\pi,3\pi/2$, independent of $e'$. Inspection of equation (\ref{Hplanesegg1}) further reveals that $\bar{\bar{\Kam}}$ is locally elliptic at $\Omega=\pi/2$ and $3\pi/2$ but is hyperbolic at $\Omega=0$ and $\pi$.

\begin{figure}[tbp]
\includegraphics[width=\columnwidth]{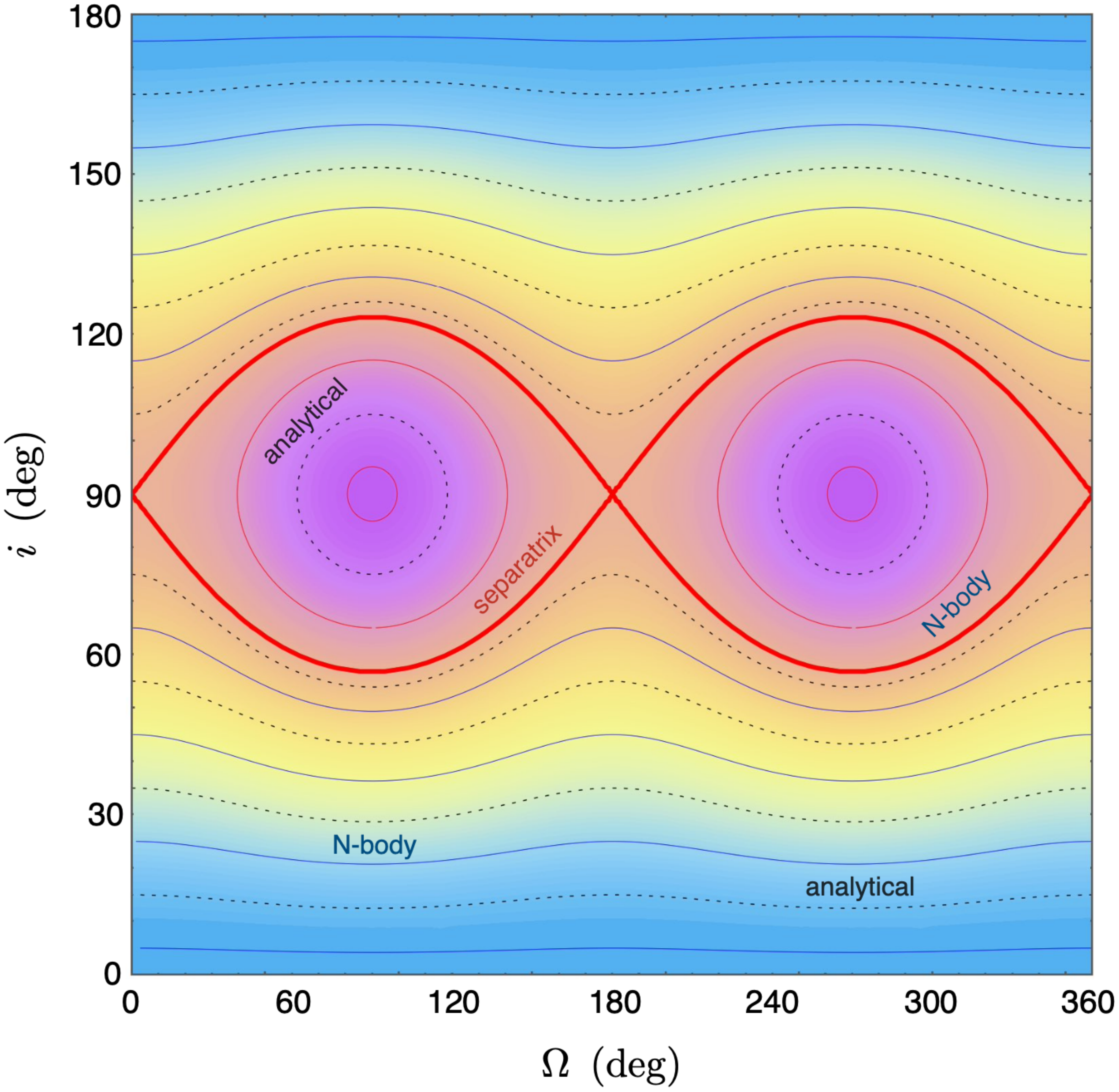}
\caption{Phase-space portrait of Hamiltonian (\ref{Hplanes}) for $e'=3$, projected onto the $(i,\Omega)$ plane. As in Figure (\ref{F:e_integ}), the background color-scale and dotted lines are obtained analytically, while the purple curves represent the results of $N$-body simulations where the test particle eccentricity is restored to zero between encounters. The phase-space diagram is characterized by a pendulum-like second-order resonant structure with equilibria corresponding to an orthogonal orbital configuration with $i=90\deg$.}
\label{F:i_integ} 
\end{figure}

The phase-space portrait of Hamiltonian (\ref{Hplanes}) for $e'=3$ is shown on Figure (\ref{F:i_integ}). Specifically, the background color-scale as well as the dotted lines represent level curves of equation (\ref{Hplanes}). Qualitatively, Hamiltonian (\ref{Hplanes}) possesses the typical structure of a mathematical pendulum i.e., retrograde and prograde circulation trajectories at $i\sim0$ and $i\sim\pi$ enclose a second-order resonance centered on $i=\pi/2$ \citep{Morbybook}. The separatrix of the resonance that partitions regions of $\Omega-$libration from circulation is emphasized with a solid red curve. 

The inclination half-width of this resonance is readily calculated by evaluating the separatrix equation at its $\Omega=\pi/2$ apex:
\begin{align}
\Delta i = \frac{\pi}{2}-\arccos\Bigg( \sqrt{\frac{2\big(e'^2-1 \big)^{3/2}}{\big(e'^2-1 \big)^{3/2} + 3\,\kappa\,e'^2}} \Bigg).
\label{reswidth}
\end{align}
Examination of this expression as a function of $e'$ illustrates that in the extreme limit of $e'\rightarrow1$, $\Delta i\rightarrow0$. Conversely, for $e'\gg1$, the resonance half-width asymptotically approaches $\Delta i \rightarrow \pi/2-\arccos(\sqrt{1/2})=\pi/4$. Indeed, unlike the case of planar encounters considered above, where the topology of the dynamical portrait changed at a critical value of $e'\approx3.59$ (Figure \ref{F:phase_space}), the qualitative features of the phase-space diagram shown in Figure (\ref{F:i_integ}) apply across all perturber eccentricities. Accordingly, to avoid redundancy, we will omit displaying a counterpart to Figure (\ref{F:phase_space}) pertinent to $i-\Omega$ dynamics.

\begin{figure}[tbp]
\centering
\includegraphics[width=0.95\columnwidth]{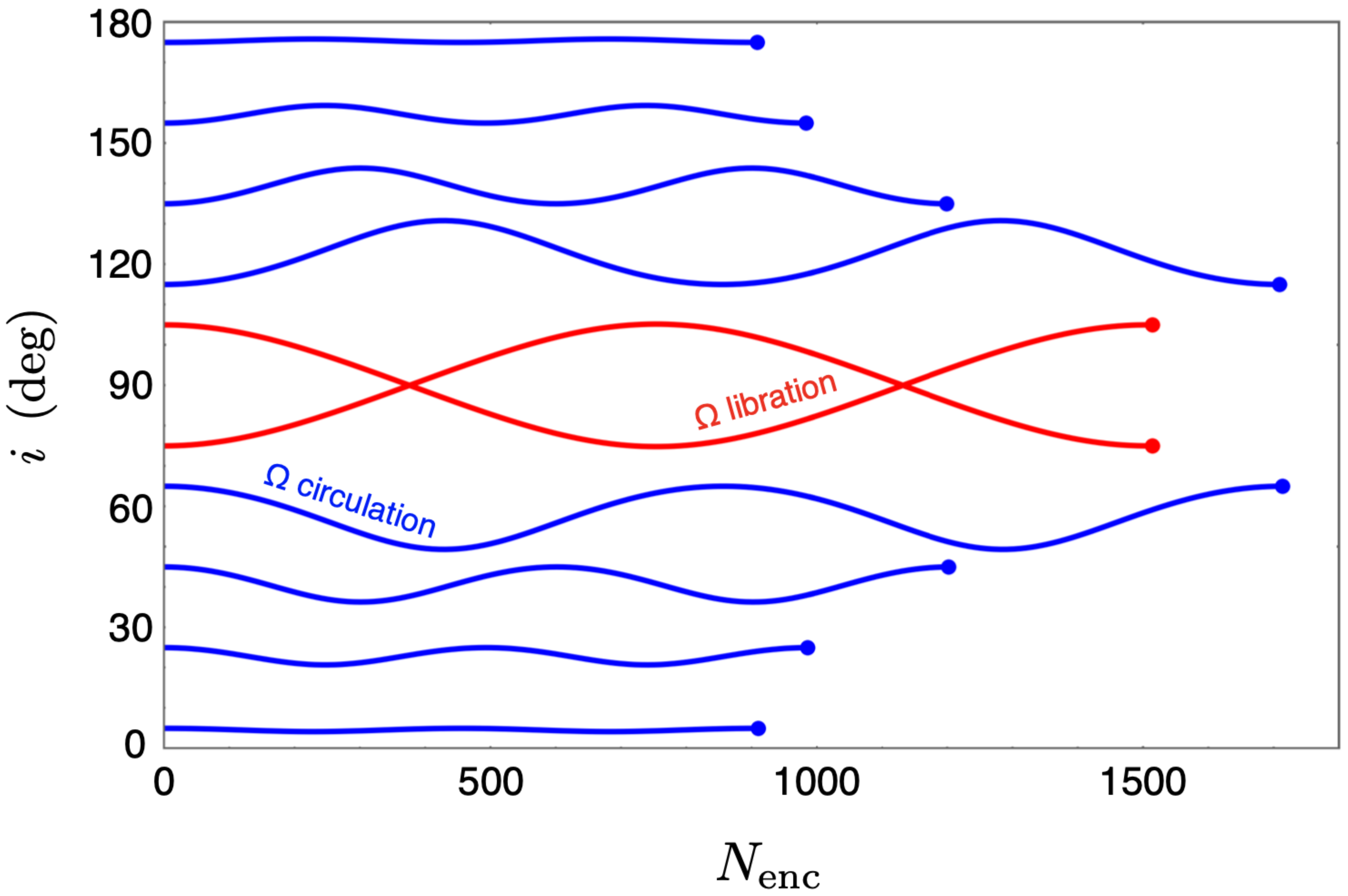}
\caption{Inclination evolution of a circular test particles with $a/b'=0.03$ under repeated encounters from a $m'=M_{\star}$, $e'=3$ perturber. Orbits entrained in a secular inclination resonance with the perturber are shown in red, while trajectories outside of the resonant domain are shown in blue. The evolution is plotted over a single circulation/libration period in $\Omega$.}
\label{F:i_enc} 
\end{figure}

As in the previous section, we can turn to the integrability of Hamiltonian (\ref{Hplanes}) to directly compare our analytic results to numerical experiments. In particular, we carried out a series of $N$-body simulations employing the same setup as above (i.e., $a/b'=0.03$, $m'/\M=1$, etc.) to recreate the level-curves of our secular model, without resorting to orbit-averaging. Notably, in order to enforce the $e=0$ limit, in these simulations we artificially restored the test particle's eccentricity back to zero after every encounter, allowing all other parameters to evolve self-consistently. The resulting $i-\Omega$ evolution computed using direct $N$-body integration over thousands of encounters is depicted in Figure (\ref{F:i_integ}) using solid purple	 lines. In light of the self-evident similarity between analytical and numerical contours depicted on the graph, we confirm the validity of our approximation scheme in the $e=0$ special case of the hyperbolic encounter problem. To complement the phase-space diagram shown in Figure (\ref{F:i_integ}), in Figure (\ref{F:i_enc}), we also show the numerically generated time-series of test particle orbital inclination, resulting from thousands of repeated encounters with $e'=3$ companions over a single circulation/libration period of $\Omega$.

\paragraph{Extension to Astrophysical Disks} With the test-particle limit of the hyperbolic encounter problem quantified, let us now consider the dynamics of a radially extended axisymmetric disk, subject to slow perturbations from passing stars. For the purposes of this work, we will limit the scope of our calculations to an idealized scenario where the internal (magneto-)hydrodynamic and self-gravitational forces of the disk are envisioned to maintain perfect coplanarity among neighboring annuli, meaning that the we will treat the disk as a rigid body. Under this assumption, every infinitesimal ring that comprises the disk has the same $i$ and $\Omega$, meaning that the Hamiltonian of the system can be obtained by averaging the system radially, weighing each annulus by its orbital angular momentum (e.g., \citealt{Bat2012}).

Let us suppose that the disk is characterized by a power-law the surface-density profile \citep{armitage}:
\begin{align}
\Sigma=\Sigma_0\bigg( \frac{a_0}{a} \bigg)^\beta,
\label{sigma}
\end{align}
where $\beta<5/2$. Then, the angular momentum stored in an annulus of radial extent $da$ is $d\Lambda=2\,\pi\,\Sigma\,a\,\sqrt{\G\,\M\,a}\,da$. Noting that all semi-major axis dependence of $\bar{\bar{\Kam}}$ is in the factor that proceeds the square brackets in equation (\ref{Hplanes}), it will be the only quantity affected by angular momentum-weighted radial averaging process. Accordingly, the pre-factor of the rigid disk Hamiltonian takes the form:
\begin{align}
&\frac{2\,\pi}{16\,a'^3\, e'^2\,(e'^2-1)^{3/2}\,n'}\frac{m'}{M} \nonumber \\
&\times\bigg(\int_{0}^{\mathcal{L}}a^3\sqrt{\frac{\G\,\M}{a^3}}\Sigma\,a\,\sqrt{\G\,\M\,a}\,da\bigg) \nonumber \\
&\times\bigg(\int_{0}^{\mathcal{L}}2\,\pi\,\Sigma\,a\,\sqrt{\G\,\M\,a}\,da\bigg)^{-1} \nonumber \\
&=\frac{5-2\,\beta}{32(4-\beta)}\sqrt{\frac{\G\,\M}{\mathcal{L}^3}}\frac{m'}{\M}\frac{(\mathcal{L}/a')^3}{e'^2\big(e'^2-1 \big)^{3/2}\,n'},
\label{energyscale}
\end{align}
where $\mathcal{L}$ is the radial extent of the disk, and we have assumed that the inner truncation radius of the disk is much smaller than $\mathcal{L}$. 

An important conclusion that expression (\ref{energyscale}) illuminates is that with the exception of an order unity reduction of the energy scale of the governing Hamiltonian, the dynamics of a rigid disk are qualitatively identical to those of a test-particle orbiting at the disk's outer edge. Although the exact magnitude of the enhancement of evolutionary timescale is dependent upon the specific index of the surface density power law, if we adopt a \citet{1963MNRAS.126..553M} type profile with $\beta=1$, we find that the energy-scale of the disk Hamiltonian is only reduced by a factor of $(2\beta-5)/(2\beta-8)=2$ when compared with a test-particle Hamiltonian evaluated at $a=\mathcal{L}$. In other words, restricted three-body problem results at $e=0$ depicted in Figures (\ref{F:i_integ}-\ref{F:i_enc}) trivially translate to the more astrophysically relevant problem of stochastic gravitational perturbations exerted upon fluid nebulae by passing stars, and we will utilize this correspondence in the next section.

\section{Early Evolution of the Solar System} \label{section:applications}

Let us now digress from academic curiosities considered in the previous section and apply the secular formalism developed above to a pair of specific examples. The first of these exercises is a direct application of the results outlined in section \ref{section:meanfield}, and addresses the evolution of the total angular momentum vector of the giant planets of the solar system, subject to the collective potential of the birth cluster. The primary result of this analysis is that even if the solar system spent $\tau \sim100\,$Myr embedded within an open cluster composed of $N\sim3000$ stars, the obliquity acquired by the sun would not exceed $\psi\lesssim1\deg$. Thus, it is very unlikely that the sun's $6-$degree spin-orbit misalignment could plausibly be attributed to the twist of the angular momentum vector ensuing from the cluster potential. 

The second example concerns a less trivial calculation of the response of the cold classical Kuiper belt to stochastic perturbations from passing stars. In particular, we apply the stochastic secular impulse formalism outlined in section \ref{section:flybys} to the outer solar system to derive limits on the birth environment of the solar system that ensue from the preservation of the cold belt's muted inclination dispersion \citep{Brown2001}. Quantitatively, this constraint translates to the solar system's stellar number density weighted cluster residence time of less than $\sim 2 \times 10^4\,$Myr/pc$^3$. Based upon our results, we further argue that the distribution of orbital inclinations within the cold classical population is largely primordial \citep{2010ApJ...722L.204P,BatKB2011,Nesvorny2019}, and stems almost exclusively from gravitational self-stirring. 

\subsection{Twist of the Solar System} \label{sec:twist} 

Consider the response of the giant planets of the solar system to phase-averaged evolution facilitated by Hamiltonian (\ref{Kozai}). For simplicity, let us adopt the $\upsilon=2$ Plummer profile and envision that the sun's orbital radius within the cluster corresponds to $\xi_{\rm{max}}$ (i.e., $a'=\sqrt{2/3}\,c$), such that our estimates yield an effective upper limit on the computed effect. In the same vein, let us recall the fiducial model cluster parameters quoted in section \ref{section:meanfield}: $M_{\infty}=1200\,M_{\odot}$, $c=0.35\,$pc, yielding $\Psi_{\rm{c}}\approx 2/3\,$(AU/year)$^2$. 

With these specifications in place, the characteristic frequency of cluster-induced perihelion precession can be obtained by setting $i\rightarrow0$ in equation (\ref{PlummerHam}) and applying Hamilton's relation
\begin{align}
\frac{d\varpi}{dt}&=\frac{-1}{\sqrt{\G\,M_{\odot}\,a}} \frac{\partial \bar{\bar{\mathcal{H}}}}{\partial\Gamma}\sim \frac{9}{25}\,\sqrt{\frac{3}{5}}\,\bigg(\frac{a}{c} \bigg)^2\frac{\Psi_{\rm{c}}}{\sqrt{\G\,M_{\odot}\,a}}.
\end{align}
For our baseline cluster parameters and $a\lesssim40\,$AU, the above expression evaluates to $d\varpi/dt\lesssim0.001$''/yr. By comparison, secular eigen-frequencies of the Lagrange-Laplace solution of the outer solar system are on the order of $g\gtrsim1\,$''/yr and thus exceed cluster-induced perihelion precession by more than three orders of magnitude\footnote{It is likely that at the early stages of the solar system's post-nebular evolution, the orbital architecture of the giant planets was more compact than it is today \citep{Tsiganis2005}, yielding even faster secular perihelion precession than that entailed by the Lagrange-Laplace solution applied to the present-day solar system.}  \citep{1950USNAO..13...81B,md1999}. As briefly mentioned in section \ref{section:meanfield}, this implies that the cluster-induced Kozai-Lidov resonance will be adiabatically suppressed by planet-planet interactions. In turn, this means that the harmonic term in equation (\ref{Kozai}) can be ignored (that is, averaged over), and the planetary eccentricities can be taken to be null. 

After these simplifications, Hamiltonian (\ref{Kozai}) reduces to:
\begin{align}
\bar{\bar{\mathcal{H}}}=-\frac{9\,\Psi_{\rm{c}}}{100}\sqrt{\frac{3}{5}}\,\bigg(\frac{a}{c} \bigg)^2\,\cos^2(i).
\label{fjdalfjdsakdf}
\end{align}
A key characteristic of this expression is that the only dynamical variable it depends on, is the inclination. Therefore, for the system at hand, the sole consequence of the birth cluster's mean field will be the nodal regression of the solar system's mean plane, as defined by the solar orbit within the cluster. 

Following the same reasoning as in section \ref{sec:inc_circ_orb}, we treat the giant planet orbits as a set of rigid rings confined to a common plane, and compute the nodal regression rate of the system by applying Hamilton's equation $d\Omega/dt=(\partial\bar{\bar{\mathcal{H}}}/\partial H)/\sqrt{\G\,M_{\odot}\,a}$ and weighting each planet's contribution by its angular momentum:
\begin{align}
\bigg\langle \frac{d\Omega}{dt} \bigg\rangle = - \frac{9}{50}\sqrt{\frac{3}{5}}\,\frac{M_{\infty}}{M_{\odot}}\,\frac{\cos(i)}{\Xi\,c^3}\,\sum_{j=5}^{8}n_j\,m_j\,a_j^{7/2},
\label{noderatemean}
\end{align}
where $\Xi=\sum_j m_j \sqrt{a_j}$. In order to evaluate this expression, we have to specify the architecture of the giant planets. In this regard, it is crucial to note that the orbits of the giant planets almost certainly experienced significant divergent migration early in the solar system's lifetime, owing to a transient dynamical instability that ensued due to their interactions with a $\sim20\,M_{\oplus}$ primordial disk of planetesimals extending from $\sim15\,$AU to Neptune's present-day orbit \citep{Tsiganis2005,nesvorny}. This means that during the epoch relevant to cluster-induced dynamics, the orbital configuration of the giant planets was likely more tightly packed than today's solar system.

\begin{figure*}[tbp]
\centering
\includegraphics[width=\textwidth]{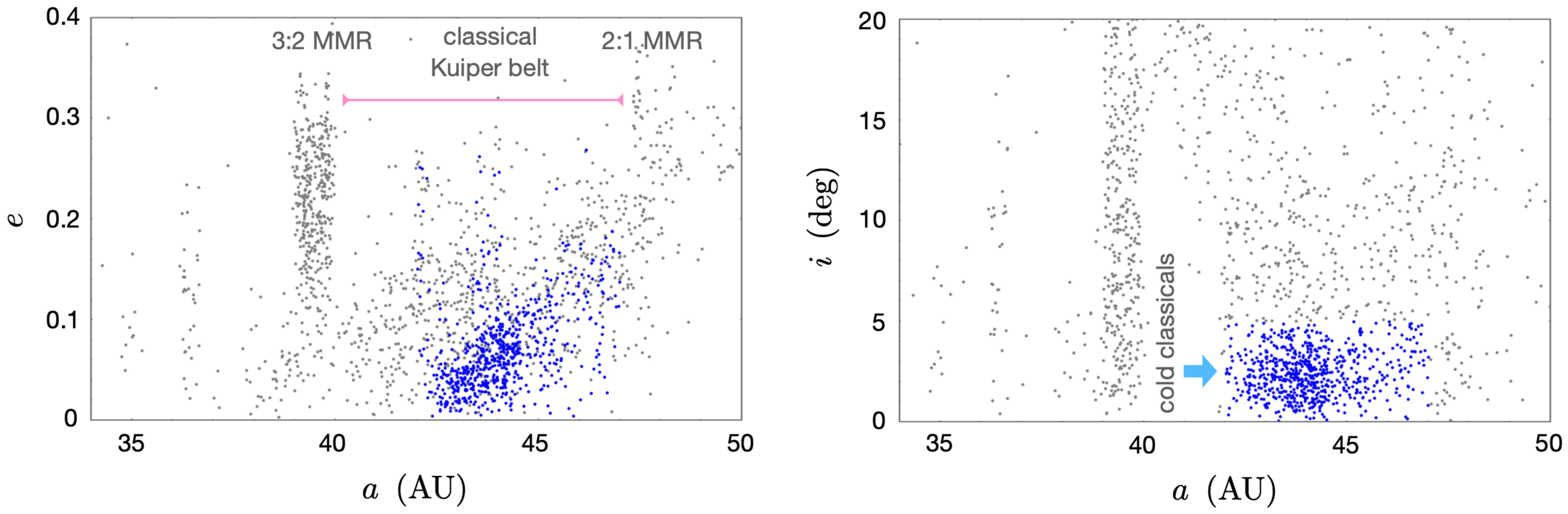}
\caption{Observational census of the classical region of the Kuiper belt. The left and right panels of the Figure show the semi-major axis -- eccentricity and the semi-major axis -- inclination distributions of detected trans-Neptunian objects. The classical Kuiper belt, primarily residing in between the exterior 3:2 and 2:1 mean motion resonances with Neptune, is sub-divided into the dynamically ``hot" and ``cold" populations. The cold belt is nominally taken to be comprised of objects with $i\leqslant5\deg$, and is highlighted on the Figure in blue. The shown data were retrieved from the Minor Planet Center database on June 1st, 2019.}
\label{F:KBO_aei} 
\end{figure*} 

The inferred existence of the Oort cloud (\citealt{oort,2019AJ....158...43K}, and the references therein) necessitates that the (Nice model) dynamical instability unfolded after the dispersal of the birth cluster. This is because the outward ejection of $\sim20\,M_{\oplus}$ of planetesimals that occurred during the instability was the last major expulsion of icy material into the trans-Neptunian region, and had this event occurred while the cluster was still present, the Oort cloud would have been rendered unbound by passing stars. Consequently, for the calculation at hand, we adopt a compact multi-resonant configuration for the giant planets where Jupiter and Saturn, as well as Uranus and Neptune are locked into 3:2 mean motion resonances while Saturn and Uranus are entrained into a 4:3 resonance, which has been previously shown to adequately serve as an initial condition for the Nice model instability (although we also note that the specific choice of resonance indexes does not affect our results on a qualitative level; \citealt{batbro2010,nesvorny}). The planetesimal disk is modeled as a series of 20 concentric rings, equally spaced between $15$ and $35\,$AU, each containing $1\,M_{\oplus}$ of material.

For our fiducial cluster parameters, and a cluster lifetime of $\tau=100\,$Myr, the total change in the node of the solar system's mean plane given by equation (\ref{noderatemean}) is a mere $\Delta\Omega=\langle\dot{\Omega}\rangle\,\tau\approx0.7\deg$ for $\cos(i)\sim1$. Translated into solar obliquity, $\psi$, we obtain an even smaller quantity. That is, if we assume that the spin-axis of the sun is not adiabatically coupled to the planets as the most optimistic scenario (see e.g., \citealt{2016AJ....152..126B}), then a twist of the solar system's mean plane necessarily results in spin-orbit misalignment, but its magnitude cannot exceed $2\,i$ in principle. It is trivial to demonstrate that solar obliquity generated by the process takes the form
\begin{align}
\psi=\arccos(1+\sin(i)(\cos(\Delta\Omega)-1))\approx\sin(i)\,\Delta\Omega.
\end{align}
Given that $\sin(i)\cos(i)\leqslant1/2$, our nominal cluster parameters yield $\psi\lesssim0.35\deg$ -- more than an order of magnitude smaller than the sun's actual $6$-degree obliquity. 

For completeness, we repeated the above calculation with the $\upsilon=1$ Hernquist profile, keeping $M_{\infty}$ and $c$ the same, but setting the dimensionless radius to a somewhat lower value of $\xi=1/2$. This choice alters the coefficient in front of Hamiltonian (\ref{fjdalfjdsakdf}) to $5/27$ -- less than a factor of $3$ larger than the Plummer value, thus only boosting the degree of stellar obliquity excited over $100\,$Myr to $\psi\approx1\deg$. To translate this estimate to even lower (a-priori improbable; \citealt{adams2010}) values of $\xi$, we note that unlike the Plummer profile, equation (\ref{HernquistHam}) shows that the Hamiltonian associated with the Hernquist profile does not have a maximum in $\xi$ and instead grows monotonically as $\sim1/\xi$ for $\xi\lesssim1$.

Cumulatively, the analysis carried out in this section indicates that the solar obliquity is very unlikely to be rooted in long-term interactions of the planetary orbits with the sun's birth cluster. While it is possible to consider alternative combinations of variables (e.g., a more massive, longer-lived open cluster) to engineer the desired result, such a solution would almost unavoidably be contrived. In other words, the procedure of simply choosing astrophysically plausible cluster parameters is unlikely to yield values of $\psi$ in excess of $\sim1\deg$.

\subsection{Heating the Cold Classical Kuiper Belt}
\label{sec:research} 

Having quantified the smooth component of the secular forcing exerted upon the solar system by the cumulative cluster potential, we now examine a less trivial, but arguably more consequential ramification of cluster-induced evolution of the outer solar system. Namely, this section will be dedicated to quantifying the extent of dynamical heating of the outer solar system generated by the integrated effect of individual stellar fly-bys. By and large, in this section, we will make use of the stochastic secular formalism outlined in section \ref{sec:inc_circ_orb}.


Among the first major results that stemmed from observational mapping of the trans-Neptunian region two decades ago \citep{1993Natur.362..730J} has been the determination that the classical Kuiper belt -- which is primarily made up of icy debris with semi-major axes in the $a\sim42-47\,$AU range\footnote{Notably, this range of semi-major axes approximately coincides with the locations of Neptune's exterior 3:2 and 2:1 mean motion resonances.} -- is comprised of two dynamically separate components: the \textit{hot}, and the \textit{cold} populations \citep{Brown2001}. The boundary between these two constituents of the classical belt is not sharp, but is nonetheless often drawn at an orbital inclination of $i\approx5\deg$, with less inclined objects classified as being dynamically cold (\citealt{Brown2001, 2008ssbn.book...43G}; Figure \ref{F:KBO_aei}). However, because orbital inclination is conventionally measured from the ecliptic plane, this oft-cited value significantly overstates the true inclination dispersion of the cold belt \citep{Brown2004}.

An additional point of considerable importance is that because classical KBOs are affected by (secular) gravitational perturbations from Neptune, the observed orbital inclinations of KBOs can be decomposed into so-called forced and free components \citep{md1999}. Qualitatively, the forced component of the inclination is a baseline quantity that arises from interactions with the giant planets, and would persist even if some dissipative force were to be applied to the cold belt. On the contrary, the free component of the inclination is fully determined by the initial conditions of the system, and is the quantity of interest for the problem at hand. To a good approximation\footnote{In this approximation, we only account for orbit-averaged gravitational coupling of the KBOs with Neptune, and only retain the components of Neptune's secular evolution corresponding to the degenerate $f_5$ (invariable plane) and $f_8$ modes of the Lagrange-Laplace solution (see \citealt{md1999}, Ch. 7 for more details).}, a cold classical KBO's (observed) complex inclination vector, $\varsigma=i\exp(\imath\,\Omega)$, can be decomposed into the free and forced elements as follows (e.g., \citealt{BatKB2011}):
\begin{align}
\varsigma_{\rm{free}}\approx\varsigma+\frac{\mathcal{B}_8}{\mathcal{B}}\mathcal{I}_{5\,8}\,e^{\imath\,\nu_5}+\frac{\mathcal{B}_8}{\mathcal{B}-f_8} \mathcal{I}_{8\,8}\,e^{\imath\,\nu_8},
\label{etafree}
\end{align}
where $\mathcal{I}_{5\,8}=2757\times10^{-5}$, $\mathcal{I}_{8\,8}=1175\times10^{-5}$, $\nu_{5}=107.1\deg$, $\nu_{8}=202.3\deg$, $f_8=-0.68''/$yr, and
\begin{align}
&\mathcal{B}=-\frac{n}{4}\sum_{j=5}^{8}\frac{m_j}{M_{\odot}}\frac{a_j}{a}b_{3/2,j}^{(1)} \ \ \ \ \  \mathcal{B}_8=\frac{n}{4}\frac{m_8}{M_{\odot}}\frac{a_8}{a}b_{3/2,8}^{(1)} \nonumber \\
&b_{3/2,j}^{(1)}=\frac{1}{\pi}\oint\frac{\cos(\psi)\,d\psi}{\big(1-2(a_j/a)\cos(\psi)+(a_j/a)^2\big)^{3/2}}
\label{coupling}
\end{align}
are the coupling coefficients of the Lagrange-Laplace secular theory \citep{1950USNAO..13...81B}.

\begin{figure}[tbp]
\centering
\includegraphics[width=\columnwidth]{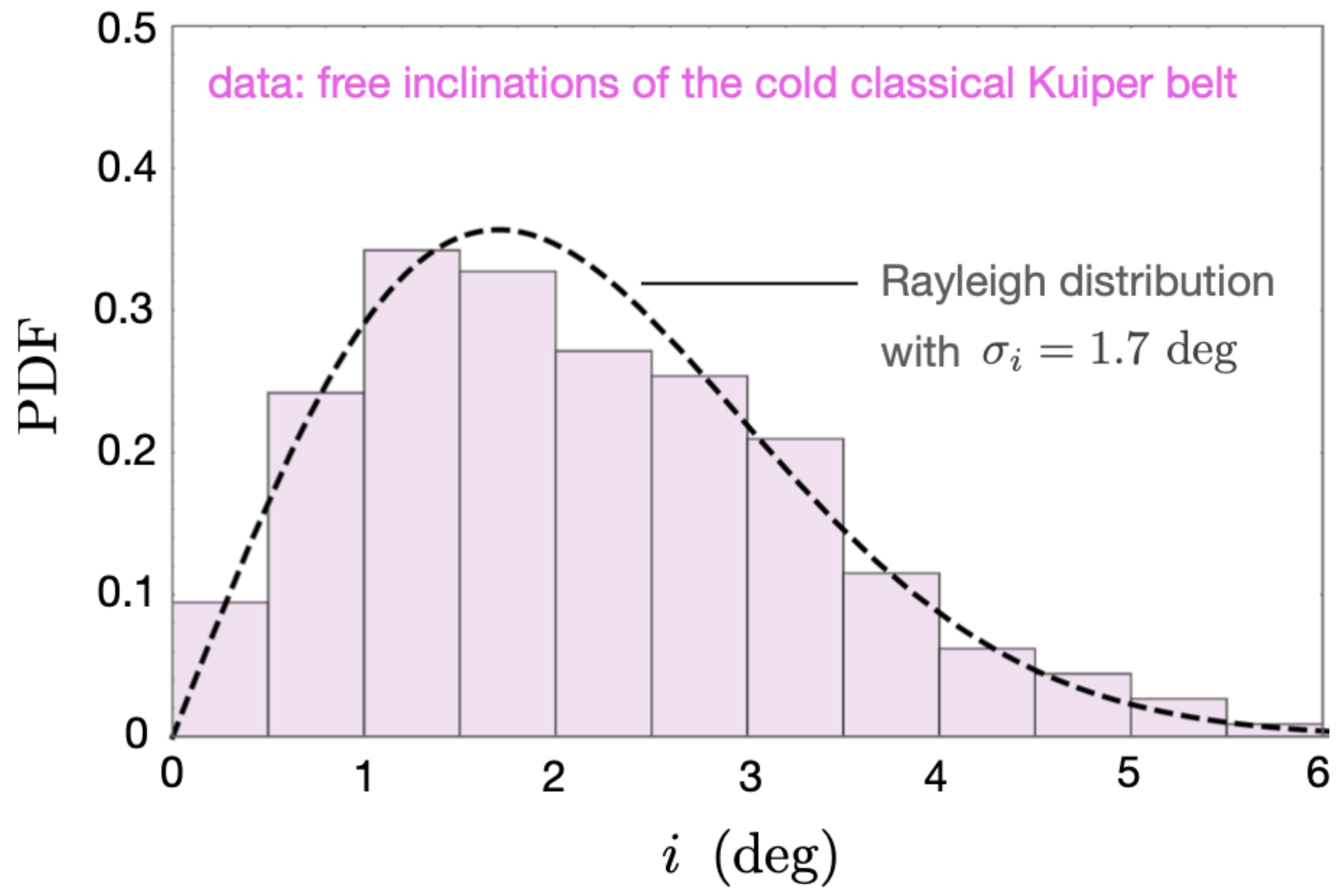}
\caption{Distribution of free inclinations of the cold classical Kuiper belt. The observational data -- shown here as a purple histogram -- is well described by a Rayleigh distribution with a scale parameter of $\sigma_i=1.7\deg$.}
\label{F:i_dist} 
\end{figure}

Figure (\ref{F:i_dist}) shows the histogram of the free inclination of the cold classical Kuiper belt. The probability density function comprised by the data are well matched by a Rayleigh distribution with a scale parameter of $\sigma_i=1.7\deg$, which is shown as with a dashed black line on the Figure. It is worth noting that by comparison, the hot classical Kuiper belt has an inclination dispersion of $\sim15\deg$ \citep{Brown2001}. Moreover, we remark that orbital eccentricities of the cold population of the classical belt are on the order of $e\sim0.05$ and are on average lower than those of the hot component, although the difference between the two populations in this degree of freedom is less dramatic (Figure \ref{F:KBO_aei}). 

Intriguingly, orbital structure comprises only one of the many characteristics in which the cold classicals appear different from the remainder of the Kuiper belt. In particular, both the (mostly red) colors and top-heavy size distribution (characterized predominantly by $\mathcal{D}\sim300\,$km objects) of cold classical KBOs are distinct from other sub-populations of the Kuiper belt 
\citep{2002ApJ...566L.125T,LykawkaMukai2005,2010Icar..210..944F}. Equally as importantly, wide binaries -- which would have been disrupted had these objects experienced close encounters with Neptune -- are present within the cold classical belt in appreciable proportion, while being markedly absent from the other classes of KBOs \citep{2010ApJ...722L.204P}. Cumulatively, these lines of evidence point towards an \textit{in-situ} formation history of the cold belt, in sharp contrast with the remainder of the Kuiper belt, which was likely dynamically emplaced from smaller heliocentric distances during the solar system's transient period of dynamical instability (\citealt{2008Icar..196..258L,BatKB2011,2012ApJ...750...43D,nesvorny2015}; see also \citealt{MorbyDavid2019} for a recent review).

\textit{If the cold classical Kuiper belt is primordial, then the maintenance of its dynamically unexcited state is a constraint that must be satisfied by the solar system's birth environment.} More generally, in light of the fact that the cold classicals may be the only population of planetesimals in the solar system that has not been significantly stirred by giant planet migration, it is of considerable interest to determine the extent of extrinsic excitation that the cold belt could have plausibly experienced, and thus illuminate the primordial inclination dispersion of planetary building blocks in the outer regions of the proto-solar nebula. 

Indeed, considerable amount of work along these lines of reasoning has already been carried out. For example, published results of numerical simulations of gravitational scattering (see e.g., \citealt{liadams15,liadams16} and the references therein) have demonstrated that the geometrical cross-section for large-scale dynamical disruption of the giant planets is $\sigma\approx2.5\times10^5\,$AU$^2$, which translates to a distance of closest approach of $r_{\rm{min}}\approx50\,$AU, where gravitational focusing is assumed to ensue with $v_{\infty}=1\,$km/s and $m'=\langle M_{\star}\rangle\approx0.4M_{\odot}$. Keeping in mind the somewhat more stringent restrictions entailed by the existence of the Kuiper belt, as a starting point of our calculations we adopt twice this value as a fiducial estimate for smallest perihelion distance, $r_{\rm{min}}\geqslant100\,$AU, that can be expected within the lifetime of the cluster. 

Because a given expectation value for the distance of closet approach can be equivalently obtained from either spending a short amount of time in a high density stellar environment or spending a long period of time in a low density stellar environment, it is convenient to define a stellar number-density-weighted residence time 
\begin{align}
\chi=\int_0^\tau n\,dt.
\label{chidef}
\end{align}
Then, the standard relationship
\be
 \pi\,r_{\rm{min}}^2\big(1+\Theta \big) \langle v\rangle \, \chi \lesssim 1,
\ee
where $\Theta=2\,\G\,(M_{\odot}+m')/r_{\rm{min}}\langle v \rangle^2$ is the Safronov number, implies $\chi \lesssim5\times10^4\,$Myr/pc$^3$ for the aforementioned crude estimate of $r_{\rm{min}}$. Limited by this product of stellar number density and cluster lifetime, let us now examine a rudimentary description of the cluster-induced evolution of a prototypical cold classical KBO from analytic as well as numeric grounds.

\paragraph{Excitation from the Plane} A simple model that can be envisioned for the early secular dynamics of the cold classical belt, is that of a single test particle located at $a=45\,$AU, evolving subject to the combined action of fixed, phase-averaged gravitational fields of the giant planets and the stochastic perturbations arising from passing stars. Within the context of this picture, giant planets force a precession of the test particle's longitudes of perihelion and ascending node with the characteristic frequency \citep{md1999}
\begin{align}
\frac{d\varpi}{dt} &\approx - \frac{d\Omega}{dt} \approx \frac{n}{4}\sum_{j=5}^8\frac{m_j}{M_{\odot}}\frac{a_j}{a}b_{3/2}^{(1)}\nonumber \\
&>\frac{3\,n}{4}\sum_{j=5}^8\frac{m_j}{M_{\odot}}\bigg(\frac{a_j}{a}\bigg)^2.
\end{align}
Referencing the results of the previous subsection, it is trivial to check that this frequency exceeds its counterpart arising from Hamiltonian (\ref{Kozai}) by a large margin, implying that the Kozai-Lidov-like mean-field dynamics of the cluster discussed in section \ref{section:meanfield} will be suppressed (see \citealt{batygin2001} for a closely related discussion). As a result, it suffices to only model the stellar fly-bys for the problem at hand.

At the same time, it is also trivial to check that in magnitude, $d\Omega/dt\ll1/\mathcal{T}_{\rm{enc}}$. The fact that this frequency is much slower than the inverse stellar crossing time means that extrinsic perturbations from passing stars will act as secular impulses that abruptly transport the KBO in phase-space on a timescale that is essentially instant compared with its usual nodal regression period. To this end, we note that if the nodal precession rate due to the giant planets greatly exceeded the rate of KBO's nodal regression induced by the star during the flyby $\sim \Delta\Omega/\mathcal{T}$, then the inclination excitation due to stellar flybys would be adiabatically suppressed, just like the Kozai-like mean-field dynamics of the cluster quoted above. As we will demonstrate below, this is not the case for the system at hand, so we do not account for secular forcing due to the giant planets \textit{during} the stellar encounters in our analytic framework for computational ease. Furthermore, we assume that the orbital eccentricity remains low enough for us to neglect all terms of order $\mathcal{O}(e^2)$ in the quadrupole-level expansion of the potential (\ref{secondorderave}). All of these simplifications will be further substantiated by direct numerical integrations that will follow.

With the above approximations in hand, we repeatedly apply the secular impulse mapping stemming from Hamiltonian (\ref{Hplanes}) to compute the inclination evolution of the test particle. The most practically straight-forward approach is to employ cartesian \Poincare\ variables \citep{Morbybook}
\begin{align}
p=\sqrt{2\mathcal{Z}}\cos{z} &&q=\sqrt{2\mathcal{Z}}\sin{z},
\label{varspq}
\end{align}
where $\mathcal{Z}=1-\cos(i)$ and $z=-\Omega$. In terms of these coordinates, the mapping equations take the form:
\begin{align}
&\Delta\,q=\frac{\partial \bar{\bar{\Kam}}}{\partial p}=\frac{p}{4}\frac{a^3}{e'^2\,b'^{3}}\frac{n}{n'}\frac{m'}{M}  \nonumber\\ 
&\times \big(3 \,\kappa\,e'^2   \left(p^2+q^2-2\right)- \left(e'^2-1\right)^{3/2} \left(p^2-2\right)\big) \nonumber \\
&\Delta\,p=-\frac{\partial \bar{\bar{\Kam}}}{\partial q}= \frac{q}{4}\frac{a^3}{e'^2\,b'^{3}}\frac{n}{n'}\frac{m'}{M} \nonumber \\
&\times\big(3 \, \kappa \, e'^2 \, \left(2-p^2-q^2\right)- \left(e'^2-1\right)^{3/2} \left(q^2-2\right) \big)
\label{impulsepq}
\end{align}

We note that the effects of individual encounters necessitate randomly drawing passing stars within the sun's immediate neighborhood in an homogeneous manner, accounting for the distribution of masses, velocities, and impact parameters. To do so, we follow the procedure outlined in \citet{Heisleretal1987} to simulate 19 distinct species of main-sequence stars with masses ranging from $\sim0.1\,M_{\odot}$ to $\sim20\,M_{\odot}$. Tables summarizing the specific stellar masses and relative number densities are provided in \citet{Heisleretal1987}. To fix an upper limit on the frequency of modeled encounters, we set the maximal impact parameter of resolved flybys to $b'_{\rm{max}}=0.1\,$pc, having checked that increasing this value does not appreciably change the results. Finally, in contrast to \citet{Heisleretal1987}, we assume a common velocity dispersion $\langle v \rangle=1\,$km/s for all stars, and draw velocities from Maxwell-Boltzmann distribution with a scale parameter\footnote{The factor of $\sqrt{2}$ arises because we are considering stellar velocity \textit{relative} to the sun, which is itself moving through the cluster.} $\sqrt{2}\,\langle v \rangle$ \citep{BT87}. This choice is motivated by observational surveys of clusters \citep{ladalada} as well as the expectation that the timescale for dynamical relaxation of the cluster is comparable to the typical lifetimes of these systems.

\begin{figure}[tbp]
\centering
\includegraphics[width=\columnwidth]{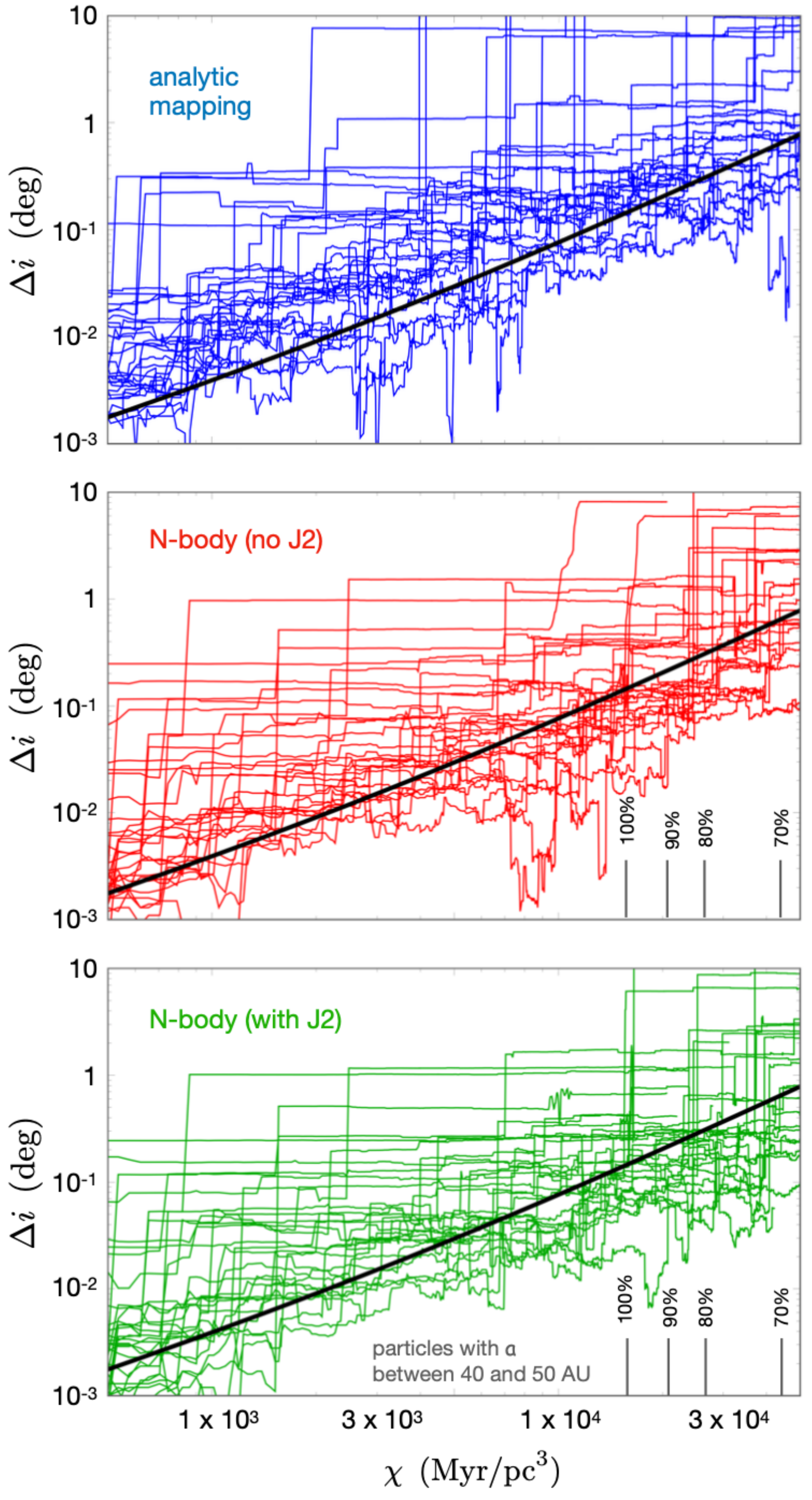}
\caption{Excitation of orbital inclination in the classical region of the Kuiper belt by stellar flybys. An initially planar test particle in orbit around the sun at $a=45\,$AU is subjected perturbations from passing stars, residing in a cluster with a velocity dispersion $\langle v\rangle=1\,$km/s. Each line represents a unique Monte-Carlo realization of the cluster environment, totaling 30 samples. The top panel depicts results computed using our analytical secular impulse model, while the middle and bottom panels show evolutions obtained through direct $N$-body integrations, with and without accounting for the phase-averaged quadrupole-level potentials of the giant planets. In panels corresponding to $N$-body simulations, the fraction of simulations where the test particles remain in the classical KBO region are labeled with large vertical ticks. The thick black lines shown in each panel correspond to the analytical inclination growth estimates given by equation (\ref{deltaieqn}).}
\label{F:deltai} 
\end{figure}

The top panel of Figure (\ref{F:deltai}) depicts the results of our analytical calculations, where the test particle was initialized at $i=0$, and subjected to perturbations arising from 30 different realizations of the cluster over a number-density-weighted timescale of $\chi=5\times10^4\,$Myr/pc$^3$. As expected, the velocity dispersion of the simulated particles grows in time, such that the average inclination at the end of the calculation is on the order of a degree. Naively, one may expect that the growth of the test particle's inclination can be understood as a diffusion-like process, wherein random perturbations from passing stars accumulate in an incoherent manner, akin to integrating over noise. As we show in the appendix (\ref{appendix:diffusion}), however, the distribution of forcings experienced by the test particle is strongly non-Gaussian and the stochastic progress of the system is always dominated by a single largest kick rather than the sum of a large number of smaller perturbations. Let us characterize this process further from purely analytical grounds.

\paragraph{An Analytic Estimate of Inclination Growth} As is well known, the characteristic rate of interactions between the solar system and passing stars can be written as $\Upsilon\sim \pi\, \eta\, b'^2\,\langle v \rangle$. The impact parameter of the closest expected approach at time $\tau$ can thus be readily derived from $\Upsilon\,\tau\sim1$. Relating the typical perturber's semi-major axis to the cluster velocity dispersion via $a'=-\G\,\mu/\langle v \rangle^2$, we obtain the minimal expected eccentricity of a perturber as a function of time:
\begin{align}
e'_{\rm{min}}\sim\sqrt{1+\frac{1}{\pi}\bigg(\frac{\langle v \rangle^2}{\G\,\mu} \bigg)^2\frac{1}{\chi\,\langle v \rangle}}.
\label{emineqn}
\end{align}
We then assume that at any value of $\chi$, the perturbations from lowest-$e'$ encounter dominates over the integrated effects of all preceding flybys (see appendix \ref{appendix:diffusion}), and simply compute the change in orbital inclination, $\Delta i$, adopting $i=0$ as an initial condition.

To account for the spherically-isotropic geometry of stellar encounters in the cluster, we express the secular impulse equations (\ref{impulsepq}) in terms of the \Poincare\ action-angle coordinates ($\mathcal{Z},z$), and average the relevant expression over the azimuthal and latitudinal angles: 
\begin{align}
\Delta i &= \sqrt{\frac{1}{4\,\pi}\int_0^2\int_0^{2\pi}\big(\Delta p^2+\Delta q^2 \big)\,dz\, d\mathcal{Z}} = \nonumber \\
&=\frac{1}{4}\frac{a^3}{e'^2\,b'^{3}}\frac{n}{n'}\frac{m'}{M}\sqrt{24\,e'^4 \,\kappa ^2+2\left(e'^2-1\right)^3}
\label{deltaieqn}
\end{align}
Substituting equation (\ref{emineqn}) into equation (\ref{deltaieqn}) thus yields the expected inclination of the cold belt, as a function of stellar number density-weighted time. The resulting curve for $\langle v\rangle=1\,$km/s and $m'=0.1\,M_{\odot}$ is shown on each panel of Figure (\ref{F:deltai}) as a black line.

Given the simplicity of the physical setup considered herein, the outlined calculation represents an additional testing ground of the secular mapping, in a realistic cluster environment. Accordingly, we repeated the performed simulations with an $N-$body model, drawing passing stars from the same distribution as above. Moreover, to assess the effect of the orbit-averaged potential of the giant planets, we carried out two sets of runs: one without a quadrupole moment, and one with a solar $J_2$ moment of magnitude
\begin{align}
J_2=\frac{1}{2}\sum_{j=5}^{8}\frac{m_j\,a_j^2}{M_{\odot}\,\mathcal{R}^2},
\label{J2eqn}
\end{align}
intended to mimic the nodal regression induced upon the test particle by Jupiter, Saturn, Uranus, and Neptune. In the latter simulation suite, we set the solar radius $\mathcal{R}=5\,$AU. The details of the simulations (integration method, etc) were identical to those carried out in section \ref{section:specialcases}. As in the analytical calculations, we only resolved encounters with an impact parameter of $b'\leqslant0.1\,$pc, and subjected the $a=45\,$AU test particle to 30 different realizations of the cluster, for $\chi=5\times10^4\,$Myr/pc$^3$.

The results of these calculations are shown in the middle panel (in red; no $J_2$) and bottom panel (in green; with $J_2$) of Figure (\ref{F:deltai}). The similarity of the test particle's evolutionary tracks depicted on the three panels of the Figure point to the fact that cluster-induced excitation of primordial planetesimals at the outer edge of the solar system is well captured by the secular mapping (\ref{impulsepq}), and that $J_2$-forced nodal regression does not appreciably suppress secular impulses facilitated by the passing stars.

An additional notion informed by the $N$-body simulations shown in Figure (\ref{F:deltai}) is that the application of the secular formalism is not sensible too far beyond $\chi\gtrsim5\times10^4\,$Myr/pc$^3$ for a  $\langle v\rangle=1\,$km/s cluster, because the probability of having an encounter that either ejects or significantly alters the specific energy of the $a=45\,$AU particle becomes appreciable. To this end, in the bottom panel of Figure (\ref{F:deltai}) we show vertical tick-marks corresponding to values of $\chi$ where the fraction of particles\footnote{The inclination evolution of particles whose semi-major axes have been altered strongly is almost always off-axis, so we simply do not plot the inclination once the semi-major axis is out of the $40-50\,$AU range.} with $a$ between $40$ and $50\,$AU equals $100\%, ... , 70\%$. Given that the chances of large-scale disruption of the Kuiper belt are approximately $\sim1/3$ at the end of the simulations, it is not worth considering greater values of $\chi$ further.


\paragraph{Inclination of the Mean Plane of the Solar System.} We carried out the preceding calculation under the assumption that the giant planets of the solar system retain a common inclination of $\langle i \rangle\approx0$ throughout the simulations. Let us now briefly verify this assumption. As already discussed in section \ref{sec:inc_circ_orb}, the dynamical response of a rigid set of orbits to external perturbations can be effectively modeled as the evolution of a single representative orbit where the accumulated changes in angular momentum are shared by the constituent wires. Accordingly, from equation (\ref{Hplanes}) it is easy to compute that angular momentum-weighted response of the four giants planets (initialized in a compact multi-resonant configuration as before) to stellar perturbations is equivalent to that of a test-wire with a semi-major axis of $a=6.4\,$AU. 


Carrying out the perturbative analysis for the giant planets, we find that over a number density-weighted timescale of $\chi=5\times10^4\,$Myr/pc$^3$, the inclination of the mean plane of the solar system is only altered by $\Delta i_{\rm{gp}}\lesssim0.1\deg$ i.e., more than an order of magnitude less than the inclination acquired by a test particle at $a=45\,$AU. This mismatch in the acquired magnitude of $\Delta i$ validates our assumption of ignoring the inclination evolution of the giant planets and modeling them as a fixed quadrupolar potential. We further note that accounting for the presence of a $\sim20\,M_{\oplus}$ planetesimal disk that extends to $30\,$AU only boosts $\Delta i_{\rm{gp}}$ by a factor of $\sim1.5$, and does not significantly alter our conclusion.

\paragraph{Inclination Distribution of Synthetic KBOs}

While the analysis carried out above demonstrates that orbital inclinations of primordial trans-Neptunian planetesimals can be excited by passing stars to the point where it becomes comparable in magnitude to the observed inclination dispersion of the cold belt, it leaves open the question of whether the resulting orbital element distribution would be compatible with the actual structure of the belt. After all, if all objects that comprise the cold classical belt traced the evolution of a single test particle exactly, the resulting distribution would simply be a $\delta-$function. In a detailed sense, however, this cannot happen because the cold classical belt spans a finite range in semi-major axis, which in turn implies differential nodal regression. Accordingly, once a finite inclination with respect to the mean plane of the solar system is acquired, each KBO would in time acquire different coordinates ($p,q$), and will thus respond to a stellar flyby in a marginally distinct manner, broadening the distribution. It is however, unclear if this process can yield a sufficiently dispersed distribution to match the real cold belt.

To answer this question quantitatively, we carried out the following elementary Monte-Carlo simulation. We began by initializing an array of 100 coplanar test particles with semi-major axes uniformly occupying the $a=42-47\,$AU range. We then subjected this group of particles to perturbations from passing stars, modifying their inclinations in accord with equations (\ref{impulsepq}). To maximize the degree of spreading of the inclination distribution, we assumed that differential nodal regression fully randomizes the longitudes of ascending nodes of the entire cold belt between encounters. As before, we continued the integration forward over a timespan corresponding to $\chi=5\times10^4\,$Myr/pc$^3$ for 30 distinct realizations of the cluster.

\begin{figure}[tbp]
\centering
\includegraphics[width=\columnwidth]{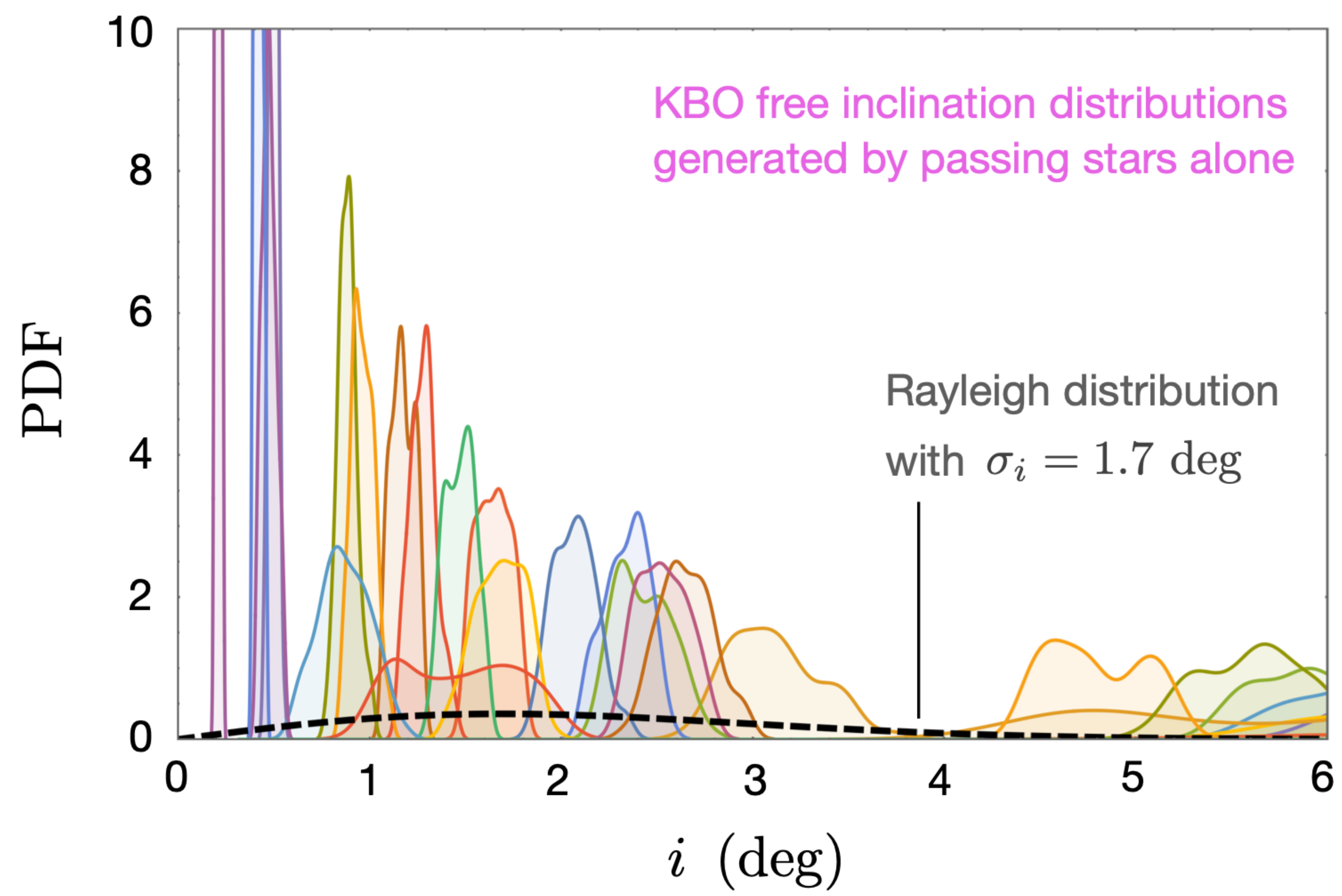}
\caption{Dispersion of orbital inclinations in the Kuiper belt, generated solely by perturbations from passing stars. Orbital distributions corresponding to discrete realizations of the solar system's birth cluster are shaded with individual colors, and the $\sigma_i=1.7\deg$ Rayleigh distribution is shown with a dashed black line for comparison. Owing to similar response to stellar flybys exhibited by all particles that comprise the model Kuiper belt, the generated distributions are much more sharply peaked than the observational data.}
\label{F:i_flyby_dist} 
\end{figure}

The probability density functions of the orbital inclinations of the generated synthetic cold belts are shown in Figure (\ref{F:i_flyby_dist}), and are shaded in different colors. The $\sigma_i=1.7\deg$ Rayleigh distribution (corresponding to the observed free inclinations of the cold belt; Figure \ref{F:i_dist}) is also shown on the Figure as a dashed black curve for comparison. Even without doing any rigorous statistical analysis, it is clear that the synthetic cold populations produced in our Monte-Carlo simulations look nothing like the actual cold belt. As anticipated above, the inclination distributions are much more sharply peaked than the observed distribution. As a result, we conclude that the inclination dispersion of the cold belt is highly unlikely to have been strongly excited by passing stars.

\begin{figure}[tbp]
\centering
\includegraphics[width=\columnwidth]{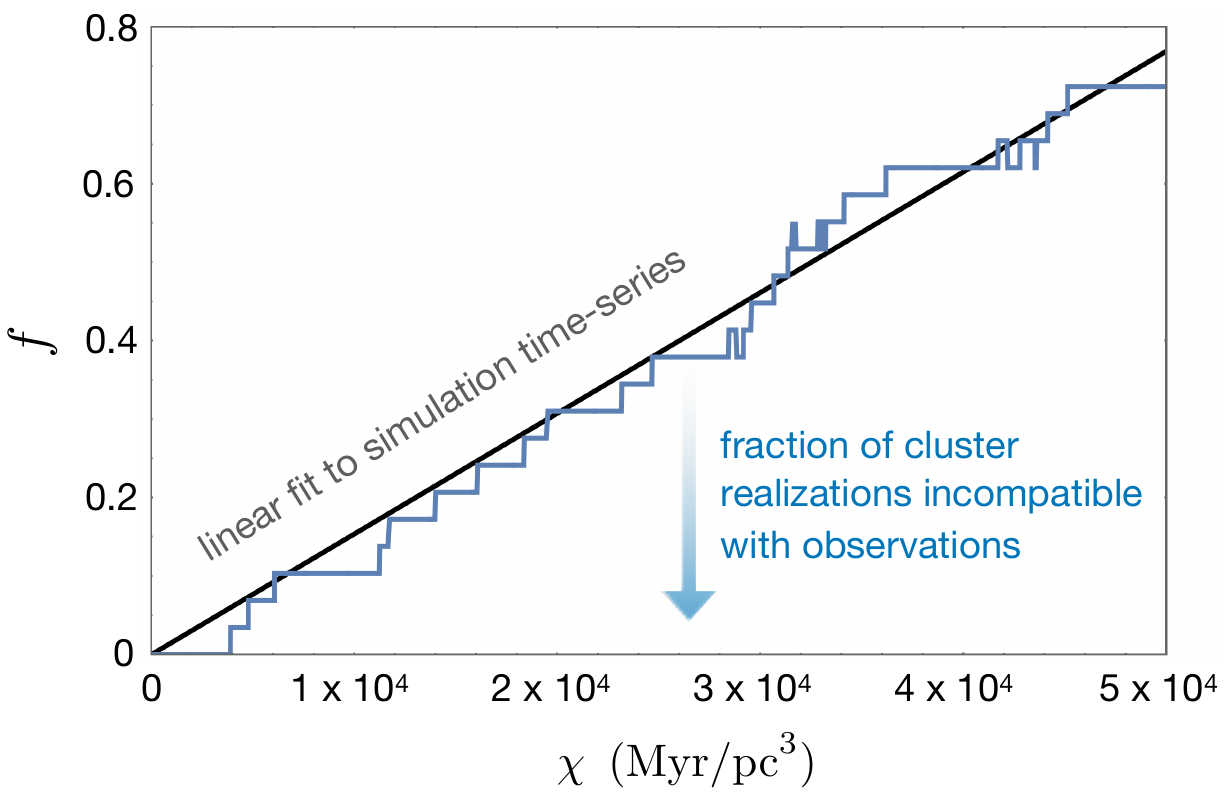}
\caption{Fraction of simulated cluster environments that are incompatible with the data, $f$, as a function of stellar number-density weighted cluster residence time, $\chi$. In these calculations, the synthetic cold classical Kuiper belt is initialized in accord with a Rayleigh distribution that adequately matches the data, yielding $f=0$ at $\chi=0$ by construction. As cluster-induced evolution of the cold belt unfolds, however, stellar encounters deform the distributions such that by $\chi=2\times10^4\,$Myr/pc$^3$, more than a quarter of the models can be rejected at the $3\,\sigma$ level. The linear fit to the simulation data given by equation (\ref{ffit}) is also shown on the Figure, with a solid black line.}
\label{F:pval} 
\end{figure}

If passing stars do not appreciably modify the orbital structure of the cold belt, and the transient dynamical instability of the giant planets tends to preserve the cold belt's dynamical architecture \citep{BatKB2011,2018ARA&A..56..137N} then it is sensible to conclude that the free inclination of the cold Kuiper belt is largely primordial in nature. In this scenario, the observed inclination distribution would be a product of gravitational self-stirring, yielding a velocity dispersion of a planetesimal disk that is comparable to the escape velocity of the planetesimals. The characteristic inclination scale is then given by the ratio of the typical escape velocity to orbital velocity. Recalling that representative cold classical KBOs have a diameter of $\mathcal{D}\sim300\,$km \citep{Nesvorny2019} and assuming a density of $\bar{\rho}=1.4\,$g/cc, this ratio evaluates to
\begin{align}
\sigma_i \sim \frac{v_{\rm{esc}}}{v_{\rm{orb}}}=\mathcal{D}\sqrt{\frac{2\,\pi\,\bar{\rho}\,a}{3\,M_{\odot}}}=1.7\deg,
\end{align}
in agreement with the observations. Moreover, the stochastic self-stirring process naturally yields Gaussian distributions of the phase-space variables ($p,q$), and noting that the Rayleigh distribution describes magnitude of a two-dimensional vector with normally distributed components, we can readily conclude that the observed inclination dispersion of the cold belt is fully compatible with a local origin, both in magnitude and distribution.

\paragraph{A Constraint on $\chi$} In light of the above results, a distinct question arises -- namely, under what conditions can the primordial architecture of the cold belt be maintained in face of cluster-induced evolution? To derive constraints on $\chi$ from the preservation of an unexcited orbital state of the cold classical population, we repeated the above Monte-Carlo experiment, this time initializing the test particles in accord with the $\sigma_i=1.7\deg$ Rayleigh distribution. As these distributions evolve forward in time within the 30 realizations of the cluster, more and more of them become incompatible with the observations. In this manner, an upper bound on the product of number density and cluster lifetime can be interpreted as the value of $\chi$ when a significant enough fraction of the simulated synthetic Kuiper belts attain an inclination dispersion that does not match that of the observations.

\begin{figure}[tbp]
\centering
\includegraphics[width=\columnwidth]{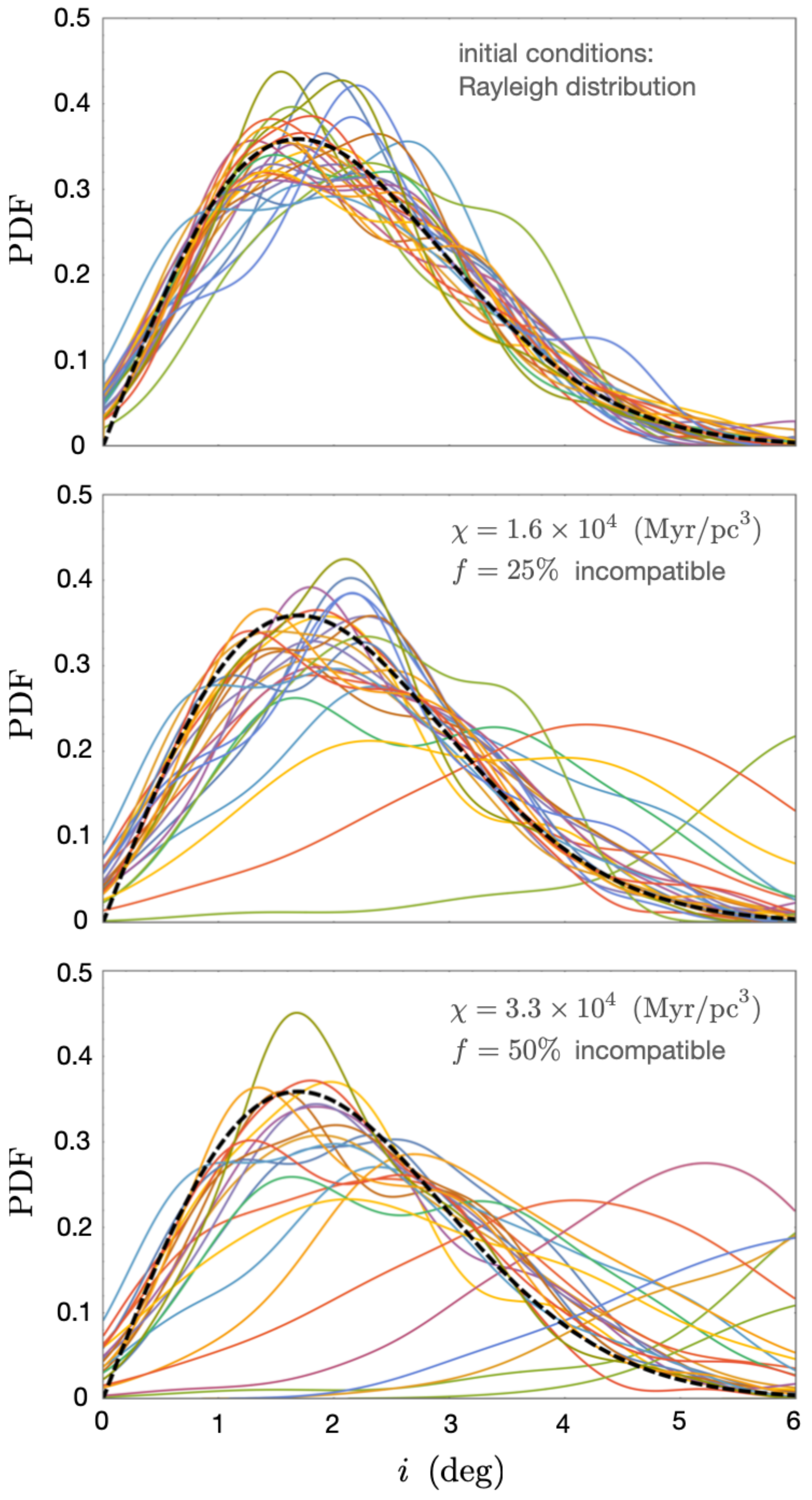}
\caption{Smoothed probability density functions of the synthetic Kuiper belts at various values of $\chi$. The top panel depicts the starting conditions, where 30 model cold classical Kuiper-belts, each composed of one hundred particles, are initialized following a Rayleigh distribution with $\sigma_i=1.7\deg$. The middle and bottom panels respectively show evolved inclination distributions, where one quarter and one half of the simulations exhibit inclination dispersions that are incompatible with the observational data.}
\label{F:dist} 
\end{figure}

As a criterion for rejection of a given distribution at a given $\chi$, we adopted a $p-$value smaller than 0.003 (i.e., 3$\,\sigma$) computed via the Kolmogorov-Smirnov test\footnote{In practice, we found that changing the critical $p-$value to either 2$\,\sigma$ or 4$\sigma$, or alternatively employing the Cram\'er-von Mises criterion instead of the KS test did not qualitatively affect our results.}. Figure (\ref{F:pval}) shows the fraction of cluster realizations, $f$, within which the simulated cold classical belt becomes incompatible with the observed one. This time-series is well matched by the approximate expression
\begin{align}
f\approx\frac{\chi}{6.5\,\mathrm{Myr/pc}^3},
\label{ffit}
\end{align}
which is shown as a black line on the Figure. Notably for $\chi\approx1.6\times10^4\,$Myr/pc$^3$ and  $\chi=3.3\times10^4\,$Myr/pc$^3$, the probability of significantly altering the orbital structure of the cold population is $\sim25\%$ and $\sim50\%$, respectively. For reference, the probability density functions of the simulated synthetic cold belts these times, as well as at $\chi=0$, are shown in Figure (\ref{F:dist}). Cumulatively, these results indicate that the upper bound on the number-density weighted lifetime of the solar system in the cluster lies at $\chi\lesssim2-3\times10^4\,$Myr/pc$^3$. 

\section{Summary} \label{summary}

The vast majority of stars -- and the planetary systems they host -- are born in young stellar associations. Dynamical interactions that ensue within these birth clusters give rise to an added degree of architectural diversity within the emergent census of planetary systems. Developing an analytical framework for quantifying the gravitational perturbations exerted upon nascent planetary systems by their birth environments, with a particular focus on the early evolution of the solar system has been the primary purpose of this work. In this concluding section, we provide a qualitative summary of the obtained results and briefly discuss their implications.

As with the current galactic environment of the sun, which affects solar system objects both via a smooth tide as well as impulsive kicks from passing stars \citep{HeislerTremaine1986,Kaib2013,Torres2019}, gravitational effects of star clusters can be subdivided into mean-field interactions and stellar fly-bys. In section \ref{section:meanfield}, we considered the former category of perturbations, and demonstrated that for a specific subset of potential-density pairs, which include the widely used \citet{Hernquist1990} and \citet{Plummer1915} models, the dynamical evolution enforced upon planetary systems by the collective potential of the cluster can be understood via a Kozai-Lidov type Hamiltonian (equation \ref{Kozai}; see also \citealt{Brasser2006,2019MNRAS.488.5489H} and the references therein). We remark that although the phase-space portrait associated with mean-field cluster interactions exhibits the usual second-order resonance in the argument of perihelion, $\omega$ \citep{KinoshitaNakai}, there exists a sizable range of parameter combinations where the circular orbit remains secularly stable even if libration of $\omega$ is possible at high eccentricity. Slow precession of the test particle's angular momentum vector, on the other hand, is an inescapable consequence of the cluster's potential.

Employing the same orbit-averaged framework, in section \ref{section:flybys} we developed a secular formalism \citep{Rasio1995,Hamers2018} for modeling perturbations arising from distant stellar fly-bys. In particular, we demonstrated that by averaging the interaction potential over the particle's orbit and integrating the resulting expression over the encounter path, we can obtain a simple Hamiltonian that adequately captures the ensuing dynamics. More specifically, this Hamiltonian contains four secular harmonics, which encapsulate three distinct physical effects: I. perturbations of the orbital planes (angular momentum vector coupling), II. hyperbolic Kozai-Lidov interactions ($e-i$ coupling), and III. prograde/retrograde apsidal eccentricity resonances (Runge-Lenz vector coupling). Comparison of our analytic results with direct $N-$body integrations across a broad range of test particle parameters and perturber eccentricities, shows agreement to better than a few percent for particle semi-major axis to perturber impact parameter ratio of $a/b'\lesssim0.1$.


The Hamiltonian describing fly-by interactions is rendered integrable in two distinctive regimes: either where the particle's orbital plane coincides with that of the perturbing star ($i=0,\pi$), or where some dissipative process (e.g., hydrodynamic interactions; \citealt{FragnerNelson2010,2014MNRAS.440.1179X}) is envisioned to consistently re-circularize the particle's orbit ($e=0$). We consider these special cases in section \ref{section:specialcases}, sequentially. In the case of planar encounters, our analysis shows that the circular orbit is stable below a critical perturber eccentricity $e'_{\rm{crit}}\approx3.59$ (for larger values it becomes a hyperbolic fixed point). This transition in the topological structure of the phase-space portrait is akin to the destabilization of the circular orbit that occurs in the context of the Kozai-Lidov resonance above a critical inclination of $i'_{\rm{crit}}\approx39\deg$ (see \citealt{Naoz2016} for a review). An interesting consequence of the existence of a critical perturber eccentricity with the orbit-averaged fly-by problem is that in 2D, the strongest encounters -- which correspond to low values of $e'$ -- are rather inconsequential for dynamically cold systems.

Our examination of the $e=0$ limit of the secular fly-by problem reveals a relatively simple picture, where the phase-space portrait of the Hamiltonian corresponds to that of a simple mathematical pendulum (see e.g. Ch. 4 of \citealt{Morbybook}). In particular, the resonance domain of this Hamiltonian is centered around an orthogonal ($i=90\deg$) orbital configuration, and the resonance width approaches $\Delta i \rightarrow0$ and $\Delta i \rightarrow45\deg$ in the $e'\rightarrow1$ and $e'\rightarrow\infty$ limits, respectively. For both, the $i=0,\pi$ and $e=0$ special cases of the secular flyby problem, we compared the analytic phase-space portraits of the governing Hamiltonian with their numeric counterparts (computed via direct $N-$body integration with $a/b'=0.035$), and found that they are essentially indistinguishable. We also considered a trivial extension of this model to account for stellar perturbations of rigid astrophysical disks and showed that radially extended structures can be modeled as test-particles residing at the outer boundaries of the disk, by reducing the effective stellar mass by a factor of order a few (e.g., exactly two for a $\Sigma\propto1/r$ \citealt{1963MNRAS.126..553M} type disk).

We applied the formalism developed in sections \ref{section:meanfield}-\ref{section:specialcases} to the solar system's early evolution in section \ref{section:applications}. We began by quantifying the integrated change in the orientation of solar system's mean plane due to the birth cluster's cumulative potential (section \ref{sec:twist}). Particular emphasis was placed on the generation of misalignment between the planetary orbits and the spin-axis of the sun, with an eye towards characterizing the cluster's contribution to the sun's present-day $6-$degree obliquity. To this end, our analysis suggests that even if the sun spent $\tau\sim100\,$Myr within a $M_{\infty}\sim1000\,M_{\odot}$ ONC-type cluster environment, the cluster-induced spin-orbit misalignment of the sun would fall short of explaining the observations by nearly an order of magnitude. While it is always possible to conjure up parameters (e.g. $\upsilon=1,\xi\ll1$) that can yield values of $\psi$ on the order of $\sim10\deg$, such configurations are a-priori unlikely and would almost certainly violate other solar system constraints \citep{adams2010}.

While our results largely rule out cluster-induced rotation of the solar system's mean plane as a viable option for excitation of solar obliquity, we note that there exist multiple other processes that are unrelated to the birth cluster, which naturally produce significant stellar obliquities. In particular, viable theories for generation of large spin-orbit misalignments during the natal disk-bearing phase of stars include magnetospheric disk-star interactions \citep{Lai2011,2015ApJ...811...82S}, disk-torquing \citep{Bat2012, 2013ApJ...778..169B, 2014ApJ...790...42S, Lai2014}, as well as asymmetric in-fall of nebular material from proto-stellar cores (\citealt{2010MNRAS.401.1505B,2015MNRAS.450.3306F}; see also \citealt{2019ApJ...879...12S} and the references therein). Moreover, observational surveys indicate that the vast majority of young embedded clusters are expected to have lifetimes of order $\tau\sim10$\,Myr, much shorter than that required to significantly affect spin-orbit alignments. As a result, in addition to applications to our solar system, our results indicate that cluster-induced evolution likely plays a negligible role in sculpting the observed distribution of spin-orbit misalignments in extrasolar planetary systems \citep{2015ARA&A..53..409W}.


In section \ref{sec:research} we carried out the second portion of our applied analysis, and considered the constraints on the solar system's birth environment emerging from the long-term preservation of the dynamically unexcited state of the cold classical population of the Kuiper belt \citep{BatKB2011,2012ApJ...750...43D,nesvorny2015}. In particular, we simulated the evolution of trans-Neptunian objects subject to perturbations from passing stars in three ways: I. using the secular impulse framework developed in section \ref{section:flybys}, II. via direct $N-$body integration of the restricted three-body problem, where stellar encounters were modeled self-consistently, and III. through $N-$body simulations of the primordial solar system where in addition to stellar flybys, quadrupolar perturbations from the giant planets were also taken into account.

Overall, we found broad quantitative agreement between all three of these approaches, implying that our analytic theory readily reproduces the results of direct $N-$body simulations at a greatly reduced computational cost, as long as stellar fly-bys are not catastrophic (such that the Kuiper belt is not destroyed).
Furthermore, we derived an almost-linear scaling of inclination growth with time, that can be understood as a tracer of the single strongest perturbation experienced by the system, rather than a diffusion-type process (see appendix \ref{appendix:diffusion}). In this vein, equation (\ref{deltaieqn}) suggests that in order for a $v_{\infty}=1\,$km/s star to disperse the Kuiper belt by $\sim1\deg$ (a value comparable to the observed inclination dispersion), an almost parabolic encounter with $e'\approx1.16$ (corresponding to an asymptotic turning angle of about $150\deg$) is required, which in turn necessitates $\chi\approx4\times10^4\,$Myr/pc$^3$. At the same time, we note that this estimate is close to the upper limit anyway, since number-density-weighted cluster lifetime itself is bounded by the fact that beyond $\chi\gtrsim5\times10^4\,$Myr/pc$^3$, encounters become sufficiently violent that the cold belt is likely to be destroyed altogether \citep{liadams15}. 

Beyond the magnitude of secular perturbations experienced by trans-Neptunian objects due to stellar fly-bys, we found that a somewhat more stringent constraint on the solar system's cluster environment can be derived by considering the spread of (free) orbital inclinations within the cold classical population. That is, while the inclination distribution of cold classicals is well-approximated by a Rayleigh distribution with a scale parameter of $\sigma_i\sim1.7\deg$, stellar encounters generate a much tighter dispersion of orbital tilts than the data, to the extent that it becomes incompatible with the observations, even if the average inclination is reproduced. In light of this disparity, we argued that the inclination dispersion of the cold classical population must be largely primordial. Indeed, a rudimentary estimate of gravitational self-stirring among $\mathcal{D}\sim300\,$km bodies within the cold belt yields an adequate explanation for the dynamical state of the cold classical population. Correspondingly, we obtained a second limit on $\chi$ by initializing the cold belt's free inclinations to follow a Rayleigh distribution with $\sigma_i\sim1.7\deg$, and demanding that stellar encounters do not alter it strongly enough to become incompatible is its starting state. Characterizing the solar system's birth environment in this way, we obtained an upper bound of number-density-weighted cluster residence time of $\chi\lesssim 2\times10^4\,$Myr/pc$^3$. Through an $n\,\sigma\,v$--type calculation, this estimate implies that in order for the cold classical Kuiper belt to have maintained its dynamically unexcited architecture, the heliocentric distance of closest approach of a passing star within the solar system's birth cluster must have been greater than $r_{\rm{min}}\gtrsim240\,$AU.



\acknowledgments We are thankful to Mike Brown, Alessandro Morbidelli, Greg Laughlin, Gongjie Li, Eduardo Marturet, Cristobal Petrovich, and Dimitri Veras for insightful discussions. We thank the anonymous referee for their careful review of the manuscript. K.B. is grateful to the David and Lucile Packard Foundation and the Alfred P. Sloan Foundation for their generous support.

\newpage

\begin{appendix}

\section{Mean-Field Dynamics: Special Cases}
\label{appendix:meanfield}

Equation (\ref{Kozai}) of the main text represents the doubly orbit-averaged interaction potential of a test particle orbiting a central body that is immersed in a spherically-symmetric background potential whose analytic form is given by equation (\ref{potential}). Recall that in these expressions, the parameter $0<\upsilon\leqslant2$ controls the sharpness of the changeover in the potential's shape across the softening length, $c$. For $\upsilon=1$, corresponding to the \citet{Hernquist1990} profile, Hamiltonian (\ref{Kozai}) can be written as follows:
\begin{align}
\bar{\bar{\Ham}}_{\upsilon=1}=-\frac{\Psi_0}{16\,(1+\xi)^3} \bigg( \frac{a}{c}\bigg)^2 \big((2+3\,e^2)((3+1/\xi)\cos^2(i)-1-3/\xi)+5\,e^2(3+1/\xi)\,\sin^2(i)\,\cos(2\omega)) \big).
\label{HernquistHam}
\end{align}
Since $\Psi_0=\G\,M_{\infty}/c$ is only a measure of the cluster's potential, it is evident that the above expression is independent of the mass of the central body, $M$, which the test particle is orbiting. This characteristic is a consequence of the implicit assumption that $M\ll M_{\infty}$, which is well satisfied for the problem of interest.

\begin{figure}[tbp]
\centering
\includegraphics[width=0.75\columnwidth]{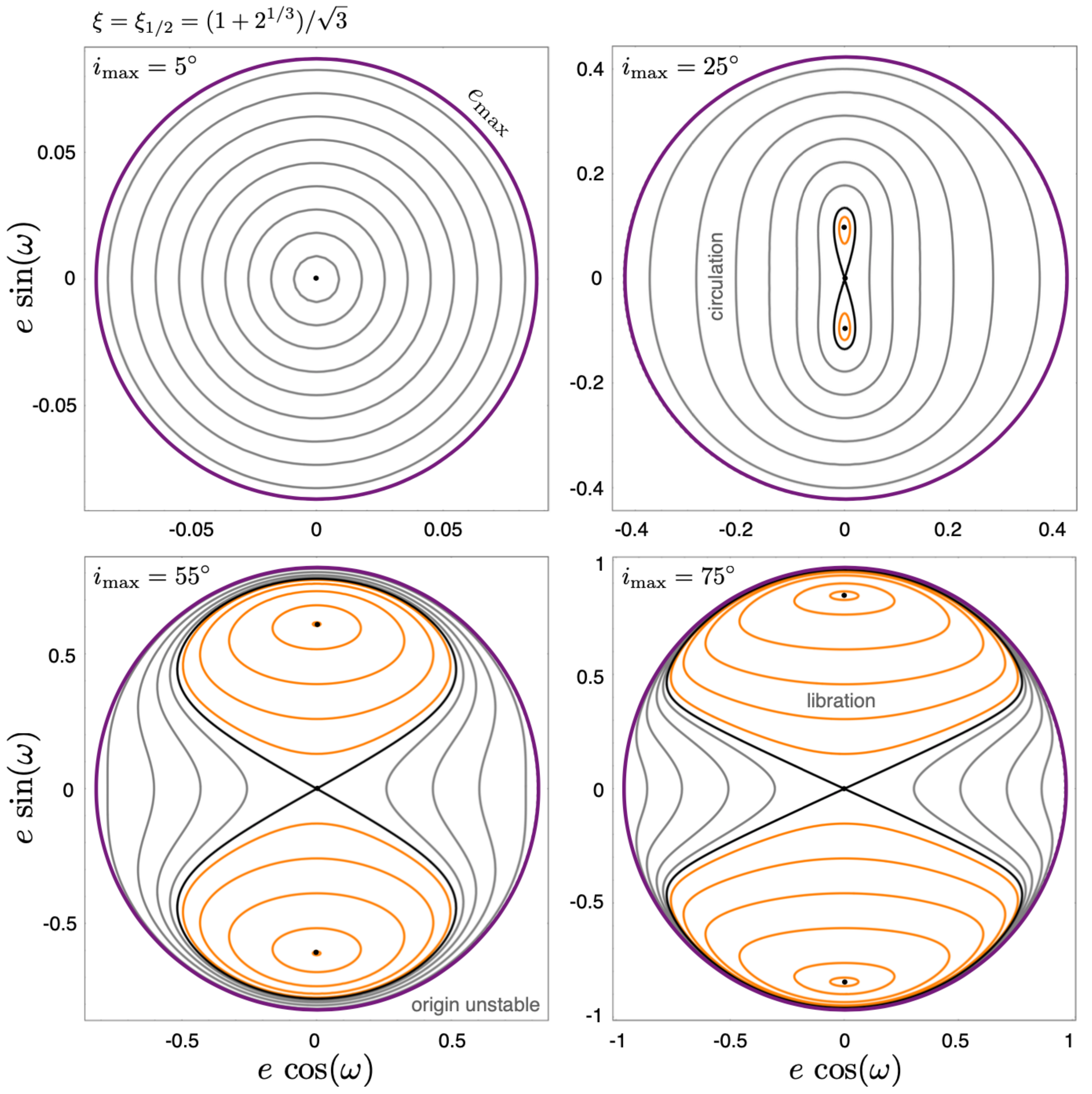}
\caption{Equivalent to Figure (\ref{F:kozai}), but for $\xi$ corresponding to the half-mass radius of the Plummer sphere. Note that unlike the $\xi<1$ case shown in Figure (\ref{F:kozai}), the $e=0$ equilibrium point becomes secularly unstable above a critical inclination in this Figure (akin to the standard Kozai-Lidov picture). The homoclinic curve running through the origin is shown with a black line.}
\label{F:Kozai_halfmass} 
\end{figure}

For the $\upsilon=2$ \citet{Plummer1915} profile, the Hamiltonian takes the form:
\begin{align}
\bar{\bar{\Ham}}_{\upsilon=2}=-\frac{\Psi_0\,\xi^2}{16\,(1+\xi^2)^{5/2}} \bigg( \frac{a}{c}\bigg)^2 \big((2+3\,e^2)(3\,\cos^2(i)-1-4/\xi)+15\,e^2\,\sin^2(i)\,\cos(2\omega)) \big).
\label{PlummerHam}
\end{align}
This expression agrees with the one given in \citet{Brasser2006} (see also the recent work of \citealt{2019MNRAS.488.5489H}). As mentioned in the main text, the pre-factor of this Hamiltonian $\propto\xi^2/(1+\xi^2)^{5/2}$ is maximized at $\xi=\sqrt{2/3}$. Conversely, in the $\xi\rightarrow\infty$ limit, both Hamiltonians (\ref{HernquistHam}) and (\ref{PlummerHam}) approach the standard Kozai-Lidov Hamiltonian \citep{KinoshitaNakai} for a test-particle perturbed by a distant mass.

To complement Figure (\ref{F:kozai}) of the main text, which shows the level curves of Hamiltonian (\ref{PlummerHam}) for $\xi\approx0.8$, in Figure (\ref{F:Kozai_halfmass}) we show an equivalent set of phase-space portraits for the dimensionless half-mass radius $\xi=(1+2^{1/3})/\sqrt{3}\approx1.3$. Here, trajectories that circulate in $\omega$ are shown in gray and ones that librate in $\omega$ are depicted in orange. Notably, the origin of the phase-space portrait already becomes hyperbolic for $i_{\rm{max}}<25\deg$ for this choice of parameters i.e., at a somewhat larger value of $\mathcal{J}$ than the standard Kozai-Lidov resonance.

\section{Collective Diffusion Versus Individual Encounters}
\label{appendix:diffusion}

In this section of the appendix, we compare the efficacy of changing the orbital elements of test particles (KBOs) due to stochastic phase space transport associated with numerous long-range stellar perturbations, and that driven by the single closest flyby. In the former case, the orbital elements change due to the accumulation of many weak (distant) encounters, and thus require a description of an ensemble of stellar kicks. Since the effects of these encounters are not correlated, the evolution can be approximately modeled as a random walk, where the total change in elements is determined by the corresponding diffusion constant.

To keep the algebraic expressions light, we consider the simple case of inclination evolution of a circular orbit in the $e'\gg1$ regime, and start with the reduced, time-integrated Hamiltonian, $\bar{\bar{\Kam}}$,  from equation (\ref{Hplanesegg1}) in the main text. To within a multiplier of order unity, the typical dimensionless step length, $\mathcal{S}$, that characterizes the random walk of the inclination angle $i$ is given by the analytic pre-factor of equation (\ref{Hplanesegg1}):
\begin{align}
\mathcal{S} \sim \frac{\alpha^3}{e'^2} \frac{n}{n'}\frac{m'}{\M} \sim \sqrt{\frac{a^3\,a'}{2\,b'^4}},
\label{SappB}
\end{align}
where we have assumed that $b'\approx a'\,e'$ and that the mass of the passing stars and the mass of the sun (or host star) are comparable, such that $m'\approx\M$. Indeed, a similar expression can be obtained directly from equation (\ref{deltaieqn}) by taking the $e'\gg1$ limit.

For small increments of the phase space variations driven by weak encounters, the changes accumulate with an effective diffusion coefficient given by 
\be
\diff = \left\langle \mathcal{S}^2 \, \Upsilon \right\rangle,
\ee
where $\Upsilon$ is the rate at which the solar system encounters other stars with impact parameter $b'$, i.e., 
\be
\Upsilon =\eta\, (\pi\, b'^2) \,\langle v\rangle.
\ee
The diffusion constant is thus given by
\begin{align}
\mathcal{D} = \int_{b'_{\rm{min}}}^{b'_{\rm{max}}} \frac{a^3\,a'}{2\,b'^4}\,\eta\, (\pi\, b'^2) \,\langle v\rangle\, \frac{2\,\pi \,b' \,db'}{ \pi\, (b'_{\rm{max}})^2} = \frac{\pi\,a^3\,a'\,\langle v\rangle\,\eta}{(b'_{\rm{max}})^2}\,\log\bigg(\frac{b'_{\rm{max}}}{b'_{\rm{min}}} \bigg),
\end{align}
where $b'_{\rm{min}}$ and $b'_{\rm{max}}$ correspond to the smallest impact parameter flyby encountered by the host star and the effective radius of the cluster, respectively.

Importantly, $b'_{\rm{min}}$ is linked to the cluster residence time by the simple relation $\Upsilon\,\tau\sim1$. Correspondingly, under the assumption of standard diffusive progress, the accrued change in inclination is given by
\begin{align}
(\Delta i)_{\rm{diff}} \sim \sqrt{\mathcal{D}\,\tau}\sim\sqrt{\frac{a^3\,a'}{(b'_{\rm{min}})^2\,(b'_{\rm{max}})^2}\log\bigg(\frac{b'_{\rm{max}}}{b'_{\rm{min}}} \bigg)}.
\end{align}
This expression can be readily compared with the change in inclination resulting from a single encounter with impact parameter $b'_{\rm{min}}$ using equation (\ref{SappB}), to give:
\begin{align}
\frac{(\Delta i)_{\rm{diff}}}{\mathcal{S}_{\rm{min}}} =\bigg(\frac{b'_{\rm{min}}}{b'_{\rm{max}}} \bigg)\sqrt{\log\bigg(\frac{b'_{\rm{max}}}{b'_{\rm{min}}}\bigg)} \ll 1\ \ \ \mathrm{for} \ b'_{\rm{max}} \gg b'_{\rm{min}}
\end{align}
The smallness of the above ratio implies that we should expect the closest encounters to dominate over the integrated effect of distant stellar perturbations.

\end{appendix}

\end{document}